\documentclass[longauth]{aa}
\usepackage{epsfig,graphicx,lscape,subfigure}
\usepackage{float,longtable,comment}
\usepackage[none,bottom,light,timestamp]{draftcopy}
\usepackage{rotating,amssymb,mathrsfs}
\usepackage[varg]{txfonts}
\usepackage{natbib}

\newcommand{\kms}{{km\,s}$^{-1}$}

\newcommand{\teff}{$T_\mathrm{eff}$\,}
\newcommand{\logg}{$\log g$\,}

\newcommand{\Msun}{\,$\rm{M}_\odot$}
\newcommand{\Lsun}{\,$\rm{L}_\odot$}
\newcommand{\Zsun}{\,$\rm{Z}_\odot$}

\newcommand{\Msp}{\,$\rm{M}_{\rm{s}}$}
\newcommand{\Mev}{\,$\rm{M}_{\rm{e}}$}
\newcommand{\Mor}{\,$\rm{M}_{\rm{o}}$}

\newcommand{\NtoH}{$\left[\frac{\rm{N}}{\rm{H}}\right]$}

\newcommand{\ve}{$v_{\rm{e}}$}
\newcommand{\vsini}{$v_{\rm{e}} \sin i$}

\newcommand{\vt}{$v_{\rm{t}}$}
\newcommand{\vor}{$v_{\rm{o}}$}
\newcommand{\vr}{$v_{\rm{r}}$}

\begin{document}

\title{The VLT-FLAMES Tarantula Survey}
\subtitle{XIX. B-type Supergiants: \\Atmospheric parameters and nitrogen abundances to investigate the role of binarity and the width of the main sequence}

\author{C. M. McEvoy\inst{1}, P. L. Dufton\inst{1}, C. J. Evans\inst{2}, V. M. Kalari\inst{3,1},  N. Markova\inst{4},  S. Sim\'{o}n-D\'{i}az\inst{5,6}, J. S. Vink\inst{3}, N. R. Walborn\inst{7}, P. A. Crowther\inst{8}, A. de Koter\inst{9,10}, S. E. de Mink\thanks{Einstein Fellow}\inst{11,12,13}, P. R. Dunstall\inst{1}, V. H\'{e}nault-Brunet\inst{14}, A. Herrero\inst{5,6}, N. Langer\inst{15}, D. J. Lennon\inst{16}, J. Ma\'{i}z Apell\'{a}niz\inst{17}, F. Najarro\inst{18}, J. Puls\inst{19}, H. Sana\inst{20}, F. R. N. Schneider\inst{15}, W. D. Taylor\inst{2}}

{\institute{Astrophysics Research Centre, School of Mathematics and Physics, Queen's University Belfast, Belfast BT7 1NN, UK
	\and {UK Astronomy Technology Centre, Royal Observatory Edinburgh, Blackford Hill, Edinburgh, EH9 3HJ, UK}
	\and{Armagh Observatory, College Hill, Armagh, BT61 9DG, Northern Ireland, UK}     
	\and{Institute of Astronomy with NAO, Bulgarian Academy of Sciences, PO Box 136, 4700 Smoljan, Bulgaria}
	\and{Instituto de Astrof\'isica de Canarias, E-38200 La Laguna, Tenerife, Spain}              
       \and{Departamento de Astrof\'isica, Universidad de La Laguna, E-38205 La Laguna, Tenerife, Spain}
 \and {Space Telescope Science Institute, 3700 San Martin Drive, Baltimore, MD 21218, USA}
     \and{Dept. of Physics \& Astronomy, Hounsfield Road, University of Sheffield, S3 7RH, UK}
\and {Astronomical Institute Anton Pannekoek, Amsterdam University, Science Park 904, 1098~XH, Amsterdam, The Netherlands}
    \and {Instituut voor Sterrenkunde, Universiteit Leuven, Celestijnenlaan 200 D, 3001, Leuven, Belgium}
\and {Astronomical Institute Anton Pannekoek, Amsterdam University, Science Park 904, 1098 XH, Amsterdam, The Netherlands}
\and{Carnegie Institution for Science: The Observatories, 813 Santa Barbara St, Pasadena, CA 91101, USA}
\and{TAPIR institute, California Institute of Technology, Pasadena, CA 91125, USA}
	\and{Department of Physics, Faculty of Engineering and Physical Sciences, University of Surrey, Guildford, GU2 7XH, UK}
	\and {Argelander-Institut f\"{u}r Astronomie der Universit\"{a}t Bonn, Auf dem H\"{u}gel 71, 53121 Bonn, Germany}
	\and{European Space Astronomy Centre (ESAC), Camino bajo del Castillo, s/n Urbanizacion Villafranca del Castillo, Villanueva de la Ca\~{n}ada, E-28692 Madrid, Spain}
	\and{Instituto de Astrof\'{i}sica de Andaluc\'{i}a-CSIC, Glorieta de la Astronom\'{i}a s/n, E-18008 Granada, Spain}
	\and{Centro de Astrobiolog\'ia (CSIC-INTA), Ctra. de Torrej\'on a Ajalvir km-4, E-28850 Torrej\'on de Ardoz, Madrid, Spain}
\and{Universit\"ats-Sternwarte, Scheinerstrasse 1, 81679 M\"unchen, Germany}
\and {ESA/STScI, 3700 San Martin Drive, Baltimore, MD 21218, USA}}

\offprints{C.M. McEvoy,\\ \email{cmcevoy14@qub.ac.uk}}

\date{Received; accepted }

\abstract{Model atmosphere analyses have been previously undertaken for both Galactic and extragalactic B-type supergiants. By contrast, little attention has been given to a comparison of the properties of single  supergiants and those that are members of multiple systems.}
{Atmospheric parameters and nitrogen abundances have been estimated for all the B-type supergiants identified in the VLT-FLAMES Tarantula survey. These include both single targets and binary candidates. The results have been analysed to investigate the role of binarity in the evolutionary history of supergiants.}
{TLUSTY non-LTE (local thermodynamic equilibrium) model atmosphere calculations have been used to determine atmospheric parameters and nitrogen abundances for 34 single and 18 binary supergiants.  Effective temperatures were deduced using the silicon balance technique, complemented by the helium ionisation in the hotter spectra. Surface gravities were estimated using Balmer line profiles and microturbulent velocities deduced using the silicon spectrum. Nitrogen abundances or upper limits were estimated from the \ion{N}{ii} spectrum. The effects of a flux contribution from an unseen secondary  were considered for the binary sample. }
{We present the first systematic study of the incidence of binarity for a sample of B-type supergiants across the theoretical terminal age main sequence (TAMS). To account for the distribution of effective temperatures  of the B-type supergiants it may be necessary to extend the TAMS to lower temperatures. This is also consistent with the derived distribution of mass discrepancies, projected rotational velocities and nitrogen abundances, provided that stars cooler than this temperature are post-red supergiant objects. For all the supergiants in the Tarantula and in a previous FLAMES survey, the majority have small projected rotational velocities. The distribution peaks at about 50 \kms with 65\% in the range 30 \kms $\le$ \vsini\ $\le$ 60 \kms. About ten per cent have larger \vsini\ ($\ge$ 100 \kms), but surprisingly these show little or no nitrogen enhancement. All the cooler supergiants have low projected rotational velocities of $\le$ 70\kms and high nitrogen abundance estimates, implying that  either bi-stability braking or evolution on a blue loop may be important. Additionally, there are a lack of cooler binaries, possibly reflecting the small sample sizes.  Single-star evolutionary models, which include rotation, can account for all of the nitrogen enhancement in both the single and binary samples. The detailed distribution of nitrogen abundances in the single and binary samples may be different, possibly reflecting differences in their evolutionary history.}
{The first comparative study of single and binary B-type supergiants has revealed that the main sequence may be significantly wider than previously assumed, extending to  \teff = 20\,000 K. Some marginal differences in single and binary atmospheric parameters and abundances have been identified, possibly implying non-standard evolution for some of the sample. This sample as a whole has implications for several aspects of our understanding of the evolutionary status of blue supergiants. }


\authorrunning{C. M. McEvoy et al.}

\titlerunning{B-type Supergiants in the Tarantula Nebula}
\maketitle
\nopagebreak[4]
%________________________________________________________________
\section{Introduction}                                         \label{s_intro}

The VLT-FLAMES Survey of massive stars \citep{eva05, eva06} provided high-quality observations of an unprecedented number of B-type stars in the Galaxy and the Large and Small Magellanic Clouds (LMC and SMC, respectively). Quantitative analysis of these objects was presented in a series of articles by \citet{duf06}, \citet{tru07}, \citet{hun07,hun08b,hun08a,hun09a}, and \citet{dun11}. 

Motivated by the desire to extend our investigation of the effects of, for example, rotational mixing and stellar winds, to more massive stars (O-type stars, B-type supergiants), 
we initiated the VLT-FLAMES Tarantula Survey \citep[VFTS,][hereafter Paper~I]{eva11} to observe  800 massive stars in the 30~Doradus region of the LMC. In the course of these observations we observed 438 B-type stars \citep{eva14}, of which approximately 10\% are luminous B-type (or late O-type)  supergiants, which are the focus of this paper.

The existence of such a large number of B-type supergiants represents a long-standing problem in stellar evolution \citep[e.g.][]{fit90}. Canonical models 
predict a gap in the Hertzsprung--Russell diagram between the main sequence O-type stars and core helium-burning blue supergiants; however, this blue Hertzsprung Gap 
is in fact filled in observational H--R diagrams. 
There is substantial uncertainty regarding the evolutionary status of these objects, even down to the most basic question of whether they are 
core hydrogen (H) burning main-sequence stars, or core helium (He) burning stars. 
Interestingly, in contradiction to hotter stars with a range of high and low rotational velocities, 
B-type supergiants with effective temperatures below 22\,000 K all seem to be slow rotators. \cite{vin10} have discussed different explanations for this, including bi-stability braking, post-red supergiant blue loops, and binary interaction.

Although B-type supergiants were once thought to be single stars \citep{hum78}, it has been suggested that binarity plays an important role in enhancing their nitrogen (N) 
abundance \citep{lan08,hun09a}. Whilst there have been several analyses over the last few decades, a comparison 
of the properties of single supergiants versus those in multiple systems has yet to be performed. 
Here we have the unique opportunity to directly compare multiple and single B-type supergiants, in a sample of approximately 50 objects (of which about a third are confirmed binaries). 
Surface N abundance has been established as one of the most important diagnostics when studying stellar evolution, since it will be 
affected by mixing, mass loss, and binarity. 

Section \ref{s_obs} gives a brief overview of the observations and data reduction, followed by an explanation of the adopted criteria for classification as a single or binary star.  Details of the model atmosphere analyses are given in Sect.~\ref{s_parameters}, and then other stellar parameters are estimated in Sect. \ref{s_st_par}. These results are discussed in Sect. \ref{s_discussion}.

\section{Observations}\label{s_obs}

The majority of the VFTS data were obtained using the Medusa mode of FLAMES, which uses fibres to feed the light from up to 130 targets simultaneously to the Giraffe spectrograph. Nine Medusa configurations (Fields~A to I) were observed in the 30~Dor region, with the targets sampling its different clusters and the local field population. The position of each of our supergiants is shown in Fig. \ref{f_positions}. Full details of target selection, observations, and data reduction were given in Paper~I.

The analysis presented here employs the FLAMES--Medusa observations obtained with two of the standard Giraffe settings (LR02 and LR03), giving coverage of the $\lambda\lambda$3960-5050\,\AA\ region at a resolving power of $\sim$7000. Each of the 540 potential B-type stars identified in Paper~I has been analysed using a cross-correlation technique to identify radial velocity variables \citep{dun15}.  Subsequent spectral classification identified more than 10\% as B-type (or O9.7-type) supergiants with luminosity classes I or II. There were  also seven additional targets with III-II luminosity classes  and these were initially included in our analysis. However all but two (VFTS\,027 and VFTS\,293) were subsequently excluded as their estimated logarithmic surface gravity (see Sect. \ref{s_parameters}) was greater than 3.2 dex, which has been used previously to delineate the end of hydrogen core burning \citep[see, for example,][]{hun08b, hun09a, hun09b}

Spectral sequences in both effective temperature and luminosity class have been given in \citet{eva14} and illustrate the high quality of the supergiant spectroscopy. Table~\ref{t_Targets} summarises  aliases, binary status (Sect. \ref{s_binarity}), spectral types \citep{eva14, wal14} and typical signal-to-noise ratios (S/Ns) for our adopted sample. S/Ns were estimated from the wavelength regions, 4200-4250 and 4720-4770\AA, which did not contain strong spectral lines. However, particularly for the higher estimates, these should be considered as lower limits as they may be affected by weak absorption lines.

 \begin{figure}
\epsfig{file=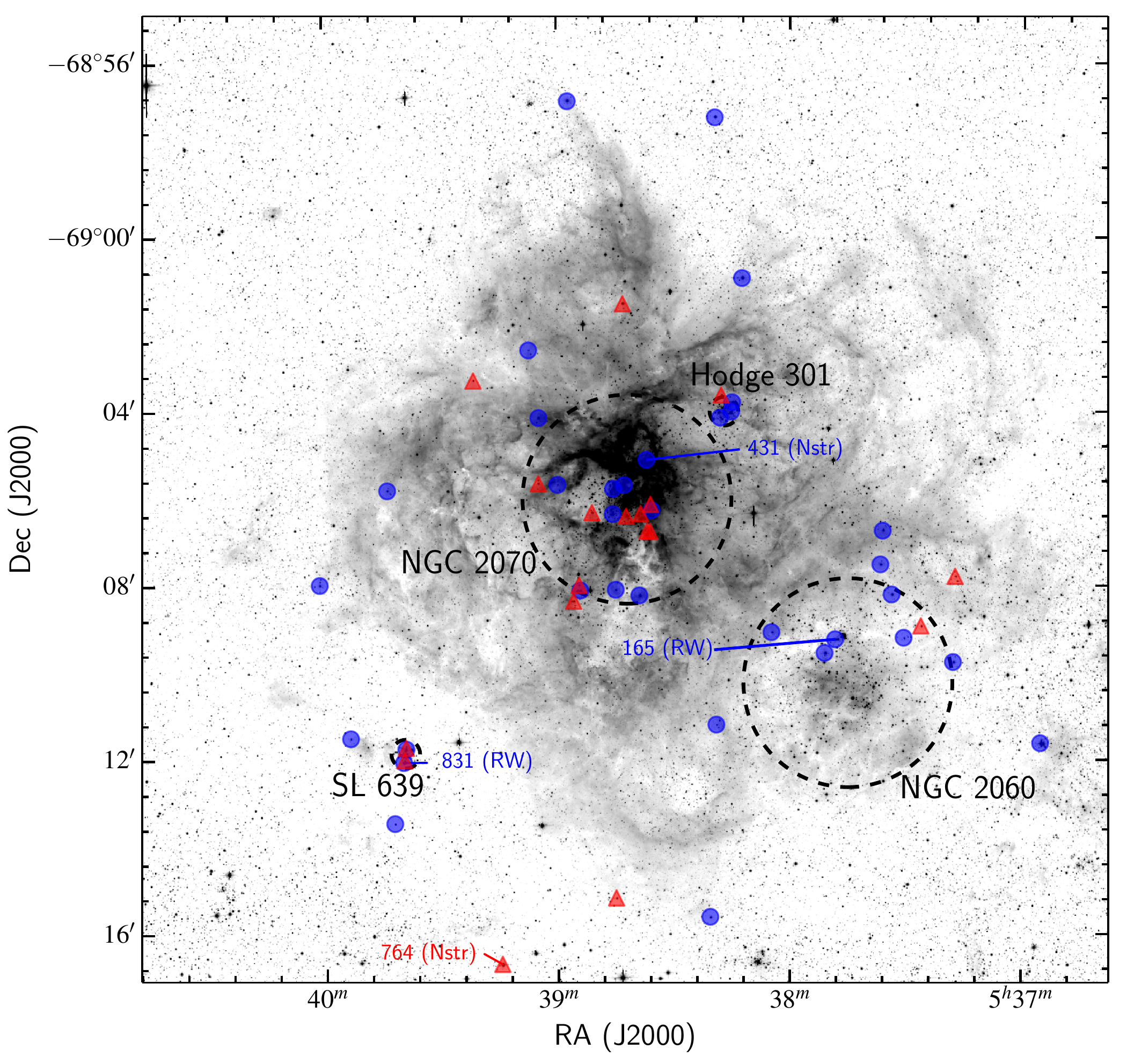,width=\linewidth,angle=0}\\
\caption{Map of positions of our sample in 30 Doradus. The binary stars are marked with red triangles and the single stars with blue circles. We have labelled four stars specifically, VFTS\,165 and 831, (candidate runaway stars) and VFTS\,431 and 764 (classified as nitrogen strong spectra).}
\label{f_positions}
\end{figure}

\begin{table*}
  \caption{Binarity status, spectral classifications \citep[normally taken from][]{eva14} and typical signal-to-noise ratios of the  VFTS B-type supergiant sample. For binarity status, D15 refers to the radial-velocity analysis of \citet{dun15}; M refers to their subsequent moment analysis and Ad is the adopted status.}\label{t_Targets}
\centering
\begin{tabular}{clrllllcl}
\hline\hline
Star & Alias & \multicolumn{4}{c}{Binarity analysis} &Spectral Type$^\dagger$ & Typical S/N \\
 &  & $\Delta$RV & D15 & M & Ad & & LR02~~LR03 & \\
\hline
003 & Sk$-$69$^\circ$~224 & 14 & S & \ldots & S & B1 Ia$^+$ & 540 ~ 420  \\
027 & \ldots & 111 &SB1 & \ldots & B & B1 III-II & 280 ~ 130 \\
028 & \ldots & 5 &S & \ldots & S & B0.7 Ia Nwk & 330 ~ 200 \\
045 & \ldots & 85 & SB1 & \dots & SB1 & O9.7 Ib-II Nwk & 100 ~ $\phantom{1}$90 \\ %ST92 1-04
060 & \ldots & 8 & S & \dots & S& B1.5 II-Ib((n)) & $\phantom{1}$90 ~ $\phantom{1}$70 \\
069 & \ldots & 2 & S & \ldots & S & B0.7 Ib-Iab & 340 ~ 270  \\ 
082 & \ldots & 2 & S & \ldots & S & B0.5 Ib-Iab & 350 ~ 310  \\
087 & \ldots & 8 & S & \ldots & S & O9.7 Ib-II & 470 ~ 330  \\
165 & \ldots & 11 & S & S & S & O9.7 Iab & 280 ~ 270 \\ 
178 & \ldots & 4 & S & \ldots & S & O9.7 Iab & 350 ~ 440 \\
232 & \ldots & 4 &S & \ldots & S & B3 Ia & 230 ~ 150  \\ %ST92 1-104
261 & \ldots & 7 & S & \ldots & S & B5 Ia & 420 ~ 310  \\
269 & R132 & 3 & S & \ldots & S & B8 Ia & 540 ~ 390 \\ %WB97-1
270 & \ldots & 5 & S & \ldots & S & B3 Ib & 290 ~ 170 \\ %WB97-2
291 & \ldots & 104 & SB1 & \ldots & SB1 & B5 II-Ib  & 180 ~ 150 \\ %WB97-8
293 & \ldots & 11 & S & \ldots & S & B2 III-II(n)e & 260 ~ 160 \\ %GC00 Be-1  
302 & \ldots & 4 & S & \ldots & S & B1.5 Ib & 160 ~ 100  \\
307 & Sk$-$68$^\circ$~136 & 6 & S & \ldots & S & B1 II-Ib & 380 ~ 270 \\
315 & \ldots & 2 & S & \ldots & S & B1 Ib & 180 ~ 110 & \\
417 & \ldots & 5 & S & \ldots & S & B2 Ib & 200 ~ 100 & \\ %S99-204
420 & Mk\,54 & 14 & S & SB1 & SB1  & B0.5 Ia Nwk & 330 ~ 290 \\ %P93-0488, S99-016
423 & Mk\,52  & 15 & S & SB1 & SB1  & B1 Ia: Nwk & 350 ~ 200 \\ %P93-0493, S99-048 
424 & R138  & 14 & S & S & S  & B9 Ia$^+$p & 540 ~ 390  \\ % S99-004 ; A0: I: [F60]
430 & P93-0538  & 121 & SB1 & \ldots & SB1  & B0.5 Ia$^+$((n)) Nwk & 102 ~ 122 \\ %S99-181 
431 & R137  & 6 & S & S & S  & B1.5 Ia Nstr & 650 ~ 420 \\ %P93-0548, S99-005; B0.5 Ia: [F60]
450 & Mk\,50 & 404 & SB2 & \ldots & SB2  & O9.7 III: +O7::  & 230 ~ 260 \\% P93-0643, S99-034
458 & Mk\,B  & 14 & S & S & S & B5 Ia$^+$p & 560 ~ 280 \\ %P93-0662 
525 & Mk\,38  & 25 & SB1 & SB1 & SB1 & B0 Ia & 350 ~ 260 \\ %P93-0930, S99-045 
533 & R142  & 4 & S & \ldots & S & B1.5 Ia$^+$p Nwk & 470 ~ 370 \\ % P93-0987, S99-003 
541 & P93-9024  & 12 & S & SB1 & SB1 & B0.5 Ia Nwk & 210 ~ 240 \\
576 & \ldots  & 76 & SB1 & \ldots & SB1& B1 Ia Nwk & $\phantom{1}$80 ~ $\phantom{1}$60 \\
578 & P93-1184  & 7 & S & \ldots & S & B1.5 Ia Nwk & 170 ~ 160  \\
590 & R141  & 15 & S & S & S & B0.7 Iab & 420 ~ 370  \\ %P93-1253, S99-009; B0.5: [F60]
591 & Mk\,12  & 21 & SB1 & S & S & B0.2 Ia & 390 ~ 370  \\ %P93-1257, S99-011 
652 & Mk\,05  & 367 & SB2 & \ldots & SB2 & B2 Ip + O9 III: $^\dagger$  & 130 ~ 220 \\
672 & P93-1661 & 11 & S & S & S & B0.7 II Nwk? & 210 ~ 160 \\
675 & P93-1674 & 60 & SB1 & \ldots & SB1 & B1 Iab Nwk & 190 ~ 130 \\
687 & P93-1737 & 57 & SB1 & \ldots & SB1 & B1.5 Ib((n)) Nwk & 180 ~ 160 \\
696 & Sk$-$68$^\circ$140 & 8 & S & \ldots & S & B0.7 Ib-Iab Nwk & 440 ~ 390  \\ 
714 & P93-1875 & 3 & S & \ldots & S & B1 Ia: Nwk & 150 ~ 110 \\
732 & Mk\,01 & 10 & S & \ldots & S & B1.5 Iap Nwk & 450 ~ 300 \\ %BI\,256, P93-1987 
733 & P93-1988 & 178 & SB1 & \ldots & SB1 & O9.7p & 120 ~200 \\
745 & \ldots & 4 & S & \ldots & S & B2.5 II-Ib & $\phantom{1}$70 ~ $\phantom{1}$60 \\
764 & Sk$-$69$^\circ$~252 & 13 & S & SB1 & SB1 & O9.7 Ia Nstr & 360 ~ 440 \\
779 & \ldots & 61 & SB1 & \ldots & SB1 & B1 II-Ib & 100 ~ 120  \\
826 & \ldots & 8 & S & \ldots & S & B1 IIn & 200 ~ 150  \\
827 & \ldots & 35 & SB1 & \ldots & SB1 & B1.5 Ib &150 ~ 90  \\
829 & \ldots & 15 & S & SB1 & SB1 & B1.5-2 II & 170 ~ 110  \\
831 & \ldots & 2 & S & \ldots & S & B5 Ia & 380 ~ 150 & \\
838 & \ldots & 15    & S & \ldots & S & B1:II(n) & 120 ~ 50 & \\
841 & \ldots & 12 & S & \ldots & S & B2.5 Ia & $\phantom{1}$50 ~ $\phantom{1}$40\\
845 & \ldots & 3 & S & \ldots & S & B1 II & 140 ~ 120 \\
855 & \ldots & 3 & S & \ldots & S & B3 Ib & 120 ~ $\phantom{1}$80 \\
867 & BI\,261 & 9 & S & \ldots & S & B1 Ib Nwk & 230 ~ 140  \\
\hline \\
\end{tabular}
\tablefoot{All O9.7 spectral types and those marked with $\dagger$ are from \citet{wal14}.}\\
\end{table*}

\subsection{Identification of binaries} \label{s_binarity}

The multiplicity analysis by \cite{dun15} adopted several criteria for identifying potential B-type binary stars. The most important was to look for statistically significant radial-velocity variations between the available epochs, for which a threshold of $\Delta$RV\,$>$\,16 km\,s$^{-1}$ was adopted. This value was an appropriate compromise between minimising the incompleteness of the detections of binarity, while ensuring that pulsations (and other sources of line-profile variations) did not lead to too many false positives. The O-type stars have been analysed by \cite{san13} using similar criteria. \footnote{VFTS 165 (O9.7 Iab) was analysed separately, and using the same methods as Dunstall et al., this star shows no significant evidence of binarity.}

Nonetheless, we still expect a number of undetected binaries with $\Delta$RV\,$<$\,16 km\,s$^{-1}$.  Indeed, on the basis of a moments analysis, five targets (VFTS\,420, 423, 541, 764, and 829) have $\Delta$RV\,$<$\,16 km\,s$^{-1}$ but show RV variations which are consistent with them being genuine binaries (see Dunstall et al.). From Fig. \ref{f_HRD}, it would appear that four of these five targets are some of the most luminous of the binaries, but have a similar distribution of surface gravities and effective temperatures and therefore would not bias the binary sample. We therefore consider these as binaries in the rest of the paper, but caution the reader that their true natures remain somewhat uncertain. Similarly, for six targets having $\Delta$RV\,$<$\,16 km\,s$^{-1}$, the variations appear to originate from pulsations, but for which we cannot rule out a binary companion; similar considerations apply to VFTS\,591 ($\Delta$RV\,$=$\,21 km\,s$^{-1}$). These seven objects are considered as single stars in the following analyses, but with similar caveats as for the binaries.

The adopted binary status for each of the supergiants is summarised in Table~1, including the two SB2 systems (VFTS\,450 and 652, Howarth et al. 2015).  In the following sections, for simplicity, we refer to the objects as either single or binary. The positions of all of the stars in the sample can be seen Fig. \ref{f_positions}. The binary and single samples are similarly distributed across the region.

\subsection{Data preparation}

The data reduction followed the procedures discussed in \citet{duf12} and only a brief description is given here. For stars with $\Delta$RV\,$\le$\,30\,\kms, the LR02 spectra from all useable exposures were combined using either a median or weighted $\sigma$-clipping algorithm. Given the relatively high S/N of the individual exposures, the final spectra from both methods were effectively indistinguishable.  The $\Delta$RV limit was based on the instrumental resolution and the intrinsic widths of the spectral features due to macroturbulent and rotational broadening. Tests using only exposures from a single epoch gave comparable results but with a lower S/N. Stars with larger RV shifts were treated in a similar manner, except that spectra from only the best LR02 epoch (in terms of S/N) plus any other epochs with similar radial velocities were combined.  This was preferable to shifting and combining exposures from different epochs, which would have exacerbated the nebular contamination.  

Normally, the LR03 exposures for the targets with $\Delta$RV\,$ >$\,30\,\kms\ were taken within a period of 3 hours and were therefore combined. The only exception was VFTS\,430 where spectra were obtained on two successive nights. Visual inspection yielded no evidence of significant velocity shifts and again all exposures were combined.

\subsection{Projected Rotational Velocities, \vsini\ }

These have been estimated from the profiles of the \ion{Si}{iii}~$\lambda$4552 or \ion{Mg}{ii}~$\lambda$4481 lines. We used the {\sc iacob-broad} tool and followed a similar strategy as described in \cite{Sim14}. In brief, this tool performs a complete line-broadening analysis including an independent computation of \vsini\ via Fourier Transform (FT) and line-profile fitting techniques (the later accounting for the effect of macroturbulent broadening). In most cases we found a very good agreement between the projected rotational velocities estimated by both approaches. For the few discrepant cases,  the FT probably gave erroneous results because of the relatively low S/N of the line and hence it was better to rely on the goodness of fit solution. For these targets, the FT estimates are indicated in brackets in Table \ref{t_Atm}.

For stars where rotation makes a significant contribution to the spectrum, we would expect that random errors on \vsini\  would be relatively small. This was confirmed by obtaining estimates from other lines in three supergiants leading to random errors of  10\% or less. However where other broadening mechanisms dominate and \vsini\ is below a certain limit ($\sim$40\,--\,50 \kms) systematic errors in the \vsini\ determination may dominate \citep[see notes in][]{Sun13,Sim14}. In these cases, \vsini\ estimates are best considered as upper limits.

We therefore recommend the following conservative error estimates:
\begin{enumerate}
\item \vsini\ $\ge 100$\ \kms: 10\%
\item $50 <$ \vsini\ $< 100$\ \kms: 20\%
\item \vsini\ $\le 50$\ \kms: lower limit: 0 \kms; upper limit: the larger of either 50 \kms or the estimate plus 20\%
\end{enumerate}

\section{Atmospheric parameters} \label{s_parameters}

\subsection{Methodology}\label{s_meth}

The analysis makes use of grids of model atmospheres calculated using the {\sc tlusty} and  {\sc synspec} codes \citep{hub88, hub95, hub98, lan07}. Details of the methods adopted can be found in \citet{hun07}, while a more detailed discussion of the grids  can be found in \citet{rya03, duf05}\footnote{See also http://star.pst.qub.ac.uk}. Here only a brief summary will be given, with further details available as discussed above.

Four model atmosphere grids have been generated with metallicities representative of the Galaxy (($\log$ ( \ion{Fe}/\ion{H})\, + 12) = 7.5 dex, and other abundances scaled accordingly), the LMC (7.2 dex), the SMC (6.8 dex), and a lower metallicity regime (6.4 dex).
Model atmospheres have  been calculated to cover a range of effective temperatures from 12\,000 K to 35\,000 K in steps not greater than 1500 K, and surface gravities ranging from near to the Eddington limit  up to 4.5 dex in steps of 0.15 dex  \citep{hun08b}.

These codes make the classical non-LTE assumptions including that the atmospheres can be considered plane parallel and that winds do not have a significant effect on the optical spectrum. Additionally, a normal helium to hydrogen ratio of 0.1 by number of atoms was assumed. The validity of such an approach was investigated by \citet{duf05}, who analysed the spectra of B-type Ia supergiants in the SMC using the current grid and also theoretical spectra generated by the {\sc fastwind} code \citep{san97,pul05}, that incorporates wind effects. \citet{duf05} found excellent agreement in the atmospheric parameters estimated from the two methods. Effective temperature estimates agreed to within  typically 500 K, logarithmic gravities to 0.1 dex and microturbulent velocities to 2 \kms. Additionally, abundance estimates agreed to typically 0.1 dex for the elements such as carbon, oxygen, silicon, and magnesium. For nitrogen, the differences were less than 0.2 dex, although there did appear to be a systematic difference of $~$0.1 dex between the two approaches. \citet{duf05} investigated this in detail and suggested that this arose from differences  in the adopted nitrogen model atoms and wind effects. 

Hence, we would expect our atmospheric parameter estimates to be reliable. Although there may be a small systematic error in our absolute nitrogen abundance estimates, relative abundance estimates should be reliable - especially given the significant range of nitrogen abundances deduced (see, for example, Table \ref{t_Single_2}). To further verify our results we have analysed some of our targets with {\sc fastwind} and the results are compared in Sect. \ref{s_comp}.

Two of the single supergiants have not been analysed. The VFTS\,838 spectra had low S/Ns, significant rotational broadening (\vsini\ $\simeq$ 240\,\kms) and no \ion{He}{ii} spectrum, preventing any estimation of its effective temperature. VFTS424 was excluded as its silicon and very weak \ion{He}{i} spectra implied that it lay beyond the low temperature limit of our grid. However, the presence of a weak \ion{N}{ii} spectrum implies that its photosphere may have a significant nitrogen enhancement. All the targets designated as binary supergiants were analysed.

The three characteristic parameters of a static stellar atmosphere (effective temperature, surface gravity, and microturbulence) are inter-dependant and so an iterative process, that assumes an appropriate LMC metallicity,  was used to estimate these values.

\subsection{Effective Temperature}

Effective temperatures (\teff) were initially constrained by the silicon ionisation balance. Equivalent widths of the  \ion{Si}{iii} multiplet (4552, 4567, 4574 \AA) were measured, together with either  the \ion{Si}{iv}  lines at 4089, 4116\AA  \,(for the hotter targets) or the   \ion{Si}{ii} lines at 4128, 4130\AA \, (for the cooler targets). In some cases it was not possible to measure the strength of either the  \ion{Si}{ii} or \ion{Si}{iv} spectrum. For these targets, upper limits were set on their equivalent widths, allowing limits to the effective temperatures to be estimated, \citep[see, for example,][ for more details]{hun07}. These normally yielded a temperature range of less than 2000 K and  their average was adopted.

For the hotter stars, independent estimates were also obtained from profile fitting the \ion{He}{ii} spectrum. The relatively low effective temperatures of the single stars led to only the \ion{He}{ii} line at 4686\AA\ being available for a significant fraction of the sample. There was excellent agreement between the  estimates using the two methods, with the largest difference being 1\,500 K (see Table \ref{t_Single_2}).

\begin{table}

\caption{Estimates for atmospheric parameters for the single stars. Effective temperature estimates are from the silicon ionisation balance (Si) or the \ion{He}{ii} profile at 4686\AA. Microturbulence estimates are from either the relative strength of the \ion{Si}{iii} lines (1) or the absolute silicon abundance (2).}\label{t_Single_2}

\begin{center}
\begin{tabular}{lcccrrcc}
\hline\hline
\vspace*{-0.25cm}\\
Star  & \multicolumn{2}{c}{\teff (K)} & \logg (cm s$^{-2}$) & \multicolumn{2}{c}{\vt \, (\kms)} \\

& Si & 4686 &  & 1  &  2      \\

\hline
003				& 21\,000 		& -			& 2.50	& 14		& 13 	\\
028				& 24\,000 		& -			& 2.75	& 14		& 14		\\
060				& 19\,500		& -			& 2.65	& 14		& 13		\\	
069				& 23\,500 		& 23\,000		& 2.75	& 15		& 15		\\
082				& 25\,500 		& 25\,500		& 3.00	& 14		& 13		\\
087				& 29\,000		& 28\,500		& 3.25	& 6		& 12		\\
165				& 27\,000		& 25\,500		& 3.00	& 6		& 14		\\
178				& 26\,500 		& 26\,500		& 3.00	& 14		& 9		\\
232				& 15\,500 		& -			& 2.30	& 8		& 8		\\	
261				& 14\,000		& -			& 2.15	& 16		& 11		\\
269				& 12\,000		& -			& 1.90	& 13		& 12		\\	
270				& 17\,500 		& -			& 2.90	& 9		& 1		\\
293				& 20\,000		& -		 	& 3.00	& -		& 0		\\
302				& 20\,500		& -			& 2.90	& 12		& 15		\\
307				& 22\,000		& -			& 2.80	& 11		& 7		\\
315				& 23\,000		& -			& 3.15	& 10		& 11		\\
417				& 18\,500		& -			& 2.55	& 12		& 10		\\
431$^*$			& 19\,000		& -			& 2.35	& 16		& 18		\\
458$^*$			& 13\,000		& -			&1.60	& 8   	& 16		\\
533$^*$                  	& 18\,000           & -                 	& 2.10      & 15        & 14    	\\
578				& 21\,500		& -			& 2.75	& 14		& 11		\\
590				& 24\,000		& 23\,000		& 2.80	& 14		& 14		\\
591				& 25\,000		&	-		&  2.80	& 13		& 14	 	\\
672				& 25\,000		& 24\,000		& 3.00	& 10		& 10		\\
696				& 23\,500		& -			& 2.75	& 13		& 11		\\
714				& 23\,500		& -			& 3.00	& 11		& 11		\\
732				& 20\,000		& -			& 2.50	& 12		& 12		\\
745				& 18\,000		& -			& 2.80	& -		& 3		\\
831				& 14\,000		& -			& 2.10	& 10		& 21		\\
841				& 17\,500		& -			& 2.40	& 13		& 20		\\
845				& 23\,500		& -			& 3.25	& 7		& 7		\\
855				& 17\,000		& -			& 2.75	& -		& 4		\\
867				& 24\,500		& 24\,000		& 3.15	& 13		& 11		\\
\hline
\end{tabular}
\end{center}
\tablefoot{$*$: atmospheric parameter estimates found from extrapolation}

\end{table}

The binary stars showed more variation between the estimates found using the \ion{He}{ii} line at 4686 \AA\ with a typical difference of ~2000 K (see Table \ref{t_Binary_2}). However this line was asymmetric for some targets, consistent with its profile possibly being affected by a wind.  The \ion{He}{ii} line at 4541 \AA\ was also analysed and yielded estimates in good agreement with those found using the silicon balance, to typically 500-1000 K. One star, VFTS 733, which is a short period binary with a large velocity amplitude had varying \ion{Si}{iii} line strengths at different epochs and so the effective temperature from the \ion{He}{ii} line at 4541\AA\ was adopted. 

The random uncertainty in our effective temperature estimates would appear to be of the order of $\pm$1000 K, or approximately 5\%, based on the high quality of the observational data and the agreement of the estimates from the helium and silicon spectra.

\begin{table}

\caption{Estimates for atmospheric parameters of the binary stars. The format is the same as for Table \ref{t_Single_2} but with an additional effective temperature estimate from a \ion{He}{ii} line (4541\AA).}\label{t_Binary_2}

\begin{center}
\begin{tabular}{lccccrr}
\hline\hline
Star  & \multicolumn{3}{c}{\teff (K)}  & \logg (cm s$^{-2}$) & \multicolumn{2}{c}{\vt (\kms)} \\
& Si & 4686 & 4541  & & 1  &  2    \\
\hline
027				&	20\,500	&	-		&	-		&	3.10	&	18	&	14		\\
045				&	29\,000	&	28\,000	&	29\,000	&	3.25	&	11	&	15		\\
291				&	13\,500	&	-		&	-		&	2.35	&	-	&	6		\\
420				&	26\,500	&	24\,000	&	26\,000	&	3.00	&	15	&	30		\\	
423				&	20\,500	&	-		&	-		&	2.50	&	14	&	15		\\
430$^*$                   &     24\,500    	&	25\,000	&	24\,000	&	2.65	&	11	&	11             \\
450				&	27\,000	&	-		&	27\,000	&	3.00	&	-	&	0		\\
525				&	27\,000	&	25\,000	&	27\,000	&	3.00	&	9	&	18		\\	
541				&	25\,000	&	23\,500	&	25\,000	&	2.90	&	13	&	15		\\	
576				&	20\,000	&	-		&	-		&	2.50	&	14	&	14		\\
652				&	21\,000	&	-		&	-		&	2.50	&	9	&	 6		\\
675				&	22\,500	&	21\,500	&	-		&	2.75	&	11	&	12		\\
687				&	20\,000	&	-		&	-		&	2.65	&	16	&	18		\\
733				&	22\,000	&	28\,500	&	29\,500	&	2.90	&	5	&	 2		\\
764				&	27\,500	&	23\,500	&	27\,000	&	3.00 & 	10 	&       18   	\\
779				&	23\,500	&	23\,500	&	-		&	3.20	&	10	&	10		\\
827$^{\dagger}$	&	21\,000	&	-		&	-		&	3.10	&	9	&	9		\\
829$^{\dagger}$	&	20\,500	&	-		&	-		&	2.90	&	17	&	13		\\
\hline
\end{tabular}
\end{center}
\tablefoot{$*$: atmospheric parameter estimates found from extrapolation\\
		$\dagger$: \teff constrained by absence of \ion{Si}{ii} and \ion{Si}{iv} spectra}
\end{table}

\subsection{Surface Gravity}
The logarithmic surface gravity (\logg) of each star was estimated by comparing theoretical and observed profiles of the hydrogen Balmer lines, H$\beta$, H$\gamma$, and H$\delta$. H$\alpha$ was not used as it was significantly affected by stellar winds. Automated procedures were developed to fit the theoretical spectra to the observed lines, with regions of best fit defined using contour maps of \logg against \teff. Using the effective temperatures deduced by the methods outlined above, the gravity could be estimated. The effects of instrumental, rotational, and macroturbulent broadening were considered for the theoretical profiles; however, these only had significant effects in the Balmer line cores where the observed profiles were badly contaminated by nebular emission. For both the single and binary samples, estimates derived from all three hydrogen lines agreed to typically 0.1 dex, which should represent a typical {\em random} uncertainty. Other systematic errors could be present owing to, for example, the uncertainty in the adopted line broadening theory or in the model atmosphere assumptions. \cite{how15} has estimated the contribution of centrifugal forces to the gravity estimates for the two supergiants VFTS\,450 and VFTS\,652. They find small corrections of $\Delta$\logg $\simeq 0.03$\,dex for these {\it relatively rapidly rotating} supergiants. Hence we have not attempted to correct our estimates in the other targets for this effect.

\subsection{Microturbulence}

Microturbulences (\vt) can be derived by removing any systematic dependance of an ion's estimated abundance on the strength of the lines of that ion. The \ion{Si}{iii} triplet (4552, 4567, and 4574\AA) was adopted as it is observed in effectively all our spectra and  as they arise from the same multiplet, the effects of  errors in the absolute oscillator strengths and non-LTE effects should be minimised.  This method has been used previously by, for example, \citet{duf05}, \citet{hun07}, and \citet{fra10}, who noted that it is sensitive to errors in the equivalent width measurements of the \ion{Si}{iii} lines. Therefore the microturbulence was also estimated from requiring that the silicon abundance was consistent with that found in the LMC. We adopted a value of 7.2 dex from \citet{hun07}, as this had been obtained using similar observational data and theoretical methods to those used here.

For one star (VFTS450, an SB2),  the maximum silicon abundance estimate (for zero microturbulence) was found to be 7.1 dex; additionally it was not possible to remove the variation of silicon abundance estimate with equivalent width.  \citet{hun07} found a similar effect for some of their targets and discussed the possible explanations in detail. Here we have adopted a zero microturbulence. For most other stars, the microturbulence estimates from the two methods agreed to within 5 \kms   \, (see Tables \ref{t_Single_2} and \ref{t_Binary_2}). Randomly changing the equivalent width estimates by 10\% (their estimated uncertainty) also led to typical variations of less than  5 \kms for both methods. This would imply that a realistic uncertainty to adopt for estimates would be of the order of $\pm$5 \kms.

\subsection{Adopted atmospheric parameters}\label{Ad_ap}

The adopted atmospheric parameters are summarised in Table \ref{t_Atm} for the single and binary stellar samples. Normally these were taken from the silicon balance for the effective temperature (`Si' in Tables  \ref{t_Single_2} and \ref{t_Binary_2}) and the requirement for a normal silicon abundance for the microturbulence (case 2). As a further test of the validity of our adopted atmospheric parameters, we have estimated magnesium abundances for all our sample using the \ion{Mg}{ii} doublet at 4481\AA. These are summarised in Table \ref{t_Atm} and are in satisfactory agreement with a LMC baseline value of 7.05 dex \citep[taken from][]{hun07}.

The two double-line spectroscopic binary supergiants included in our sample (VFTS450 and VFTS652) have been analysed in detail by  \citet{how15}, who adopted similar methods to those used here. To ensure consistency, we list our own estimates in Tables  \ref{t_Binary_2} and \ref{t_Atm} but we note that these are similar to those found by \citet{how15}, with the assumption that the supergiant supplies all the observed radiative flux.

VFTS\,826 has a relatively low effective temperature (with no observable \ion{He}{ii} spectrum) and a projected rotational velocity of over 200 \kms (with no observable \ion{Si}{ii}  and \ion{Si}{iv} features) and hence it was not possible to estimate its effective temperature. In Fig. \ref{826}, we plot its spectrum together with that for VFTS\,845 that has been convolved with a rotational broadening function appropriate to that for VFTS\,826. The two spectra are very similar confirming the similar classification. Therefore for VFTS\,826, we have simply adopted the atmospheric parameters for VFTS\,845, although the uncertainties will, of course be larger. For example, the relatively large magnesium abundance estimate would imply that our effective temperature estimate is too high or our microturbulent velocity estimate is too low.

\begin{table*}

\caption{Adopted estimates for the projected rotational velocity (\vsini, with FT estimates in brackets for relatively low S/N cases), atmospheric parameters and nitrogen abundances. For convenience the spectral types are repeated from Table \ref{t_Targets}, whilst mean radial velocities (\vr) are taken from \citet{eva14} and \citet{san13} . The single stars are listed first, followed by the binaries. Abundances are presented on the scale $\log$(X/H) + 12. $\log\mathscr{L}/\mathscr{L}\odot$ is proportional to the Eddington factor, where $\mathscr{L} = T_{eff}^4/g$ and has a maximum value of 4.6  dex (corresponding to the Eddington limit) for a normal helium to hydrogen ratio \citep{lan14}. }\label{t_Atm}

\begin{center}
\begin{tabular}{llllllllllc}
\hline\hline
Star &  Spectral Type & \vr	& \vsini	& \teff  &	 \logg&	 \vt &  \ion{Mg} & \multicolumn{2}{c}{ \ion{N} \ estimates}  & $\log\mathscr{L}/\mathscr{L}\odot$
 \\
	&& \kms & \kms &	K & cm s$^{-2}$	& \kms	&	&	$\lambda$3995  & all lines  \\	\hline
003	& B1 Ia$^+$			& 265	& 48	& 21\,000		&2.50	&13	&7.03	&	8.05		&7.98$\pm$0.10 (6)			&	4.18	\\
028	& B0.7 Ia Nwk			& 274	& 50	& 24\,000		&2.75	&14	&7.00	&	7.28		&7.32$\pm$0.06 (6)		&	4.16	\\
060	& B1.5 II-Ib((n))			& 284	&129& 19\,500		&2.65	&14	&7.20 	&	$\le$6.9	&	-							&	3.90	\\
069	& B0.7 Ib-Iab			& 281	& 46	& 23\,500		&2.75	&15	&7.08	&	7.96		&7.93$\pm$0.10 (7)		&	4.13	\\
082	& B0.5 Ib-Iab			& 283	& 49	& 25\,500		&3.00	&13	&7.15	&	7.92		&7.89$\pm$0.10 (7)				&	4.02	\\
087	& O9.7 Ib-II			& 275	& 60	& 29\,000		&3.25	&12	&7.07	&	7.29		&	-					&	3.99	\\
165	& O9.7 Iab			& 204	& 65	& 27\,000		&3.00	&14	&	-	&	7.83		&7.86$\pm$0.12 (5)   		&	4.12	\\
178	& O9.7 Iab			& 288	& 55	& 26\,500		&3.00	&9	&7.02	&	7.37		&7.49$\pm$0.16 (5)			&	4.09	\\
232	& B3 Ia				& 289	&  {42} (8)	& 15\,500		&2.30	&8	&6.90	&7.67	&7.61$\pm$0.17 (8)						&	3.85	\\
261	& B5 Ia       			& 285	& 39	& 14\,000		&2.15	&11	&7.00	&	7.57		&7.58$\pm$0.16 (7)				&	3.83	\\
269	& B8 Ia      			& 267	& 26	& 12\,000		&1.90	&12	&6.69	&	7.56		&7.75$\pm$0.16 (4)				&	3.81	\\
270	& B3 Ib	    			& 265	& 36	& 17\,500		&2.90	&1	&7.00	&	7.53		&7.57$\pm$0.17 (7)				&	3.47	\\
293	& B2 III-II(n)e			& 260	&125& 20\,000		&3.00	&0	&6.89	&	$\le$6.9	&	-								&	3.60	\\
302	& B1.5 Ib  				& 265	& 34	& 20\,500		&2.90	&15	&6.97	&	7.61		&7.69$\pm$0.08 (7)				&	3.74	\\
307	& B1 II-Ib				& 273	& 32	& 22\,000		&2.80	&7	&6.92	&	8.14		&7.90$\pm$0.19 (8)				&	3.96	\\
315	& B1 Ib				& 262	& 31	& 23\,000		&3.15	&11	&7.16	&	7.37		&7.39$\pm$0.15 (7)					&	3.69	\\
417	& B2 Ib				& 261	& 52	& 18\,500		&2.55	&10	&7.09	&	7.49		&7.53$\pm$0.12 (7)					&	3.91	\\
431	& B1.5 Ia Nstr			& 270	& 41	& 19\,000		&2.35	&18	&7.02	&	7.99		&7.98$\pm$0.10 (7)		&	4.16	\\
458	& B5 Ia$^+$p			& 276	& 33	& 13\,000		&1.60	&16	&7.05	&	8.01		&8.06$\pm$0.16 (7)		&	4.25	\\
533	& B1.5 Ia$^+$p Nwk		& 255	& 57	& 18\,000		&2.10	&14	&7.08	&	7.42		&7.39$\pm$0.04 (6)      &	4.31	 \\
578	& B1.5 Ia Nwk			& 275	& 37	& 21\,500		&2.75	&11	&7.13	&	6.96		&7.07$\pm$0.14 (5)		&	3.97	\\
590	& B0.7 Iab			& 254	& 60	& 24\,000		&2.80	&14	&7.00	&	8.00		&7.89$\pm$0.13 (7)			&	4.11	\\
591	& B0.2 Ia				& 287	& 48	& 25\,000		&2.80	&14	&6.96	&	7.38		&7.45$\pm$0.17 (3)		&	4.18	\\
672	& B0.7 II Nwk?			& 280	& 54	& 25\,000		&3.00	&10	&7.04	&	6.85		&	-							&	3.98	\\
696	& B0.7 Ib-Iab Nwk		& 266	& 53	& 23\,500		&2.75	&11	&7.03	&	7.19		&7.25$\pm$0.17 (5)		&	4.13	\\
714	& B1 Ia: Nwk			& 283	& {38} (2)	& 23\,500		&3.00	&11	&7.17	&7.18		&7.16$\pm$0.14 (4)			&	3.88	\\
732	& B1.5 Iap Nwk			& 267	& 45	& 20\,000		&2.50	&12	&7.00	&	6.91		&6.96$\pm$0.08 (4)		&	4.10	\\
745	& B2.5 II-Ib			& 276	& 33	& 18\,000		&2.80	&3	&6.76	&	$\le$7.2	&-								&	3.61	\\
826 &  B1 IIn				& 242	&222& 23\,500		&3.25		&7		&7.22 & 6.80 & -                                    &	3.63	\\
831	& B5 Ia				& 210	& 41	& 14\,000		&2.10	&21	&6.89	&	7.94		&8.02$\pm$0.16 (7)					&	3.88	\\
841	& B2.5 Ia				& 258	& {63} (28)	& 17\,500		&2.40	&20	&7.03	&	7.80		&7.79$\pm$0.15 (6)			&	3.97	\\
845	& B1 II				& 259	& {< 25} (14)	& 23\,500		&3.25	&7	&6.96	&	7.50		&7.54$\pm$0.12 (7)			&	3.63	\\
855	& B3 Ib				& 248	& {< 20} (3)	& 17\,000		&2.75	&4	&6.97	&	7.73		&7.71$\pm$0.20 (7)			&	3.56	\\
867	& B1 Ib Nwk			& 276	& 32	& 24\,500		&3.15	&11	&7.07	&	6.93		&7.16$\pm$0.20 (5)					&	3.80	\\
\hline
027	& B1 III-II				& 343	& 91$^2$ & 20\,500		&3.10	&18	&6.94	&	6.86		&6.81$\pm$0.07 (2)			&	3.54	\\
045	& O9.7 Ib-II Nwk		& 233	& 69	& 29\,000		&3.25	&15	&-		&	$\le$7.8	&-		               	&	3.99	\\
291 	& B5 II-Ib			& 202& $\leq$20	& 13\,500		&2.35	&6	&6.58	&	7.92		&-								&	3.56	\\
420	& B0.5 Ia Nwk			& 273	& 73	& 26\,500		&3.00	&30	&6.90	&	$\le$7.3	&-						&	4.09	\\
423	& B1 Ia: Nwk			& 278	& 42	& 20\,500		&2.50	&15	&7.12	&	7.04		&7.15$\pm$0.09 (6)		&	4.14	\\
430 	& B0.5 Ia+((n)) Nwk		& 208	& 98	& 24\,500		&2.65	&11	&7.47	&      $\le$7.7	&-					&	4.30	\\
450	& O9.7 III: + O7::		& 249$^1$	&123& 27\,000		&3.00	&0	&6.94	&	7.49		&7.50$\pm$0.01 (2)		&	4.12	\\
525	& B0 Ia				& 293	& 78	& 27\,000		&3.00	&18	&6.93	&	$\le$7.3	&-							&	4.12	\\
541	& B0.5 Ia Nwk			& 270	& 47	& 25\,000		&2.90	&15	&6.93	&	$\le$7.1	&-								&	4.08	\\
576	& B1 Ia Nwk			& 286	& 52	& 20\,000		&2.50	&14	&7.04	&	$\le$7.1	&-							&	4.10	\\
652	& B2 Ip + O9 III: 	& 255$^{1}$	& 72	& 21\,000		&2.50	&6	&6.85	&	8.09		&8.29$\pm$0.24 (7)		&	4.18	\\
675	& B1 Iab Nwk			& 266	& 56	& 22\,500		&2.75	&12	&7.08	&	6.93		&7.10$\pm$0.18 (4)		&	4.05	\\
687	& B1.5 Ib((n)) Nwk		& 314	&123& 20\,000		&2.65	&18	&7.00	&	6.94		&7.08$\pm$0.15 (4)					&	3.95	\\
733	& O9.7p		& 278	& 51 & 29\,500		&3.25	&5	&7.05	&	8.29		&8.27$\pm$0.18 (6)					&	4.02	\\
764	& O9.7 Ia Nstr			& 261	& 58 & 27\,500		&3.00	&18	&7.04	&	7.81		&7.79$\pm$0.23 (3)		&	4.15	\\
779	& B1 II-Ib				& 248	& 47 & 23\,500		&3.20	&10	&7.07	&	7.39		&7.45$\pm$0.13 (6)				&	3.68	\\
827	& B1.5 Ib				& 243	& 52 & 21\,000		&3.10	&9	&7.07	&	7.10		&7.14$\pm$0.06 (2)				&	3.58	\\
829	& B1.5-2 II			& 261	&138& 20\,500		&2.90	&13	&7.04	&	$\le$7.1 	&-										&	3.74	\\
\hline\hline
\end{tabular}
\end{center}
\tablefoot{{1: }{systematic velocity from orbital solution, see \cite{how15}}\\
{2: }{taken from \citet{duf12}}}
\end{table*}

\subsection{Nitrogen Abundances}

Nitrogen abundances were estimated primarily by measuring the equivalent width of the singlet transition at 3995\AA\ as this feature is one of the strongest \ion{N}{ii} lines in the optical spectrum and appears unblended. As such it was well observed in most of our sample and the estimates are summarised in Table \ref{t_Atm}. For the six binary stars where this feature was not visible, an upper limit on its possible equivalent width was estimated from an artificial absorption line. These had been broadened to agree with the widths of the rest of the metal lines and then degraded to the observed S/N. 

Other \ion{N}{ii} lines were present, including the triplet transitions between 4601-4643\AA\ and  the singlet line at 4447\AA. These lines were more prone to blending and were less consistently visible in our spectra. For each star the average of all the estimates (including the line at 3995\AA) is presented in Table \ref{t_Atm}, together with the standard deviation. The differences in the two estimates for a given star are generally consistent with these standard deviations. 

These abundance estimates will also be affected by errors in the atmospheric parameters which has been discussed by, for example, \citet{hun07} and \citet{fra10}. Using a similar methodology, we conservatively estimate a typical uncertainty of  0.2-0.3 dex. We note that this does not include systematic errors due to limitations in the adopted models.

\begin{table*}
\caption{Effects of 20\% secondary contribution. Abundances are presented on the scale log(X/H) + 12. }\label{t_Secondary_Contrib}
\begin{center}
\begin{tabular}{llllllrcc}
\hline\hline
Star &Secondary &\teff & logg & \vt &  \ion{Mg} & \multicolumn{2}{c}{\ion{N}\ estimates } 
\\
	&	Contribution	&	&	&	&	&		$\lambda$3995  & all lines \\
\hline	

687	& 	0\%	& 	20\,000	&	2.65		&	18	&	7.00 	&	6.94		&	7.00\\
&		20\% &	20\,500	&	2.75		&	26	&	7.08	&	7.02		&	7.17	\\
\hline
675	& 	0\%	& 	22\,500	&	2.75		&	12	&	7.08	&	6.93		&	7.10\\
&		20\% &	24\,000	&	3.15		&	20	&	7.11	&	7.00		&	7.20	\\
\hline
672	&	0\%	& 	25\,000	&	3.00		&	10	&	7.04 	&	6.85		&	-	\\
&		20\% &	27\,000	&	3.60		&	18	&	7.07	&	6.90		&	-	\\

\hline
\end{tabular}
\end{center}
~\\

\end{table*}

\subsection{Contribution of secondary stars}\label{secondary}

In the previous analysis we have assumed that the observed spectrum arises solely from the supergiant. For the binary targets (and potentially for the single stars that are actually part of a binary system), some of the flux will be from the secondary. For two targets (VFTS\,450 and 652), there is evidence for secondary spectra, which have been discussed by Howarth et al. 2015. For the remaining targets, there is no spectral evidence for a secondary leading to the SB1 classifications in Table \ref{t_Targets}.

For these targets, we have investigated the effects on our analysis by assuming that the secondary contributes 20\% of the light via a featureless continuum. This value was chosen because for larger contributions we might expect the secondary spectrum to become visible. If the companion is a particularly fast rotator, then it is possible that a contribution of more that 20\% would remain hidden, but for the purposes of this investigation we will assume 20\% to be a worst case scenario. A simple continuum was adopted for convenience and also because structure in the secondary spectrum (e.g. in the Balmer lines) would normally mitigate its influence. 

Three stars (two binaries and one single) with a range of effective temperatures were chosen as representative cases. The equivalent widths and hydrogen profiles of these stars were rescaled to allow for an unseen secondary in a similar manner to that used to correct for Be-type disc contamination in \citet{dun11}. Both the original atmospheric parameters and those implied by a 20\% secondary contribution are summarised  in Table \ref{t_Secondary_Contrib}. These imply that the nitrogen abundances calculated for the binary stars may be  underestimated by  0.1-0.2 dex. The implication of such an offset in nitrogen abundance estimates between the single and binary sample is considered in Sect. \ref{s_discussion}.

\begin{figure}

\epsfig{file=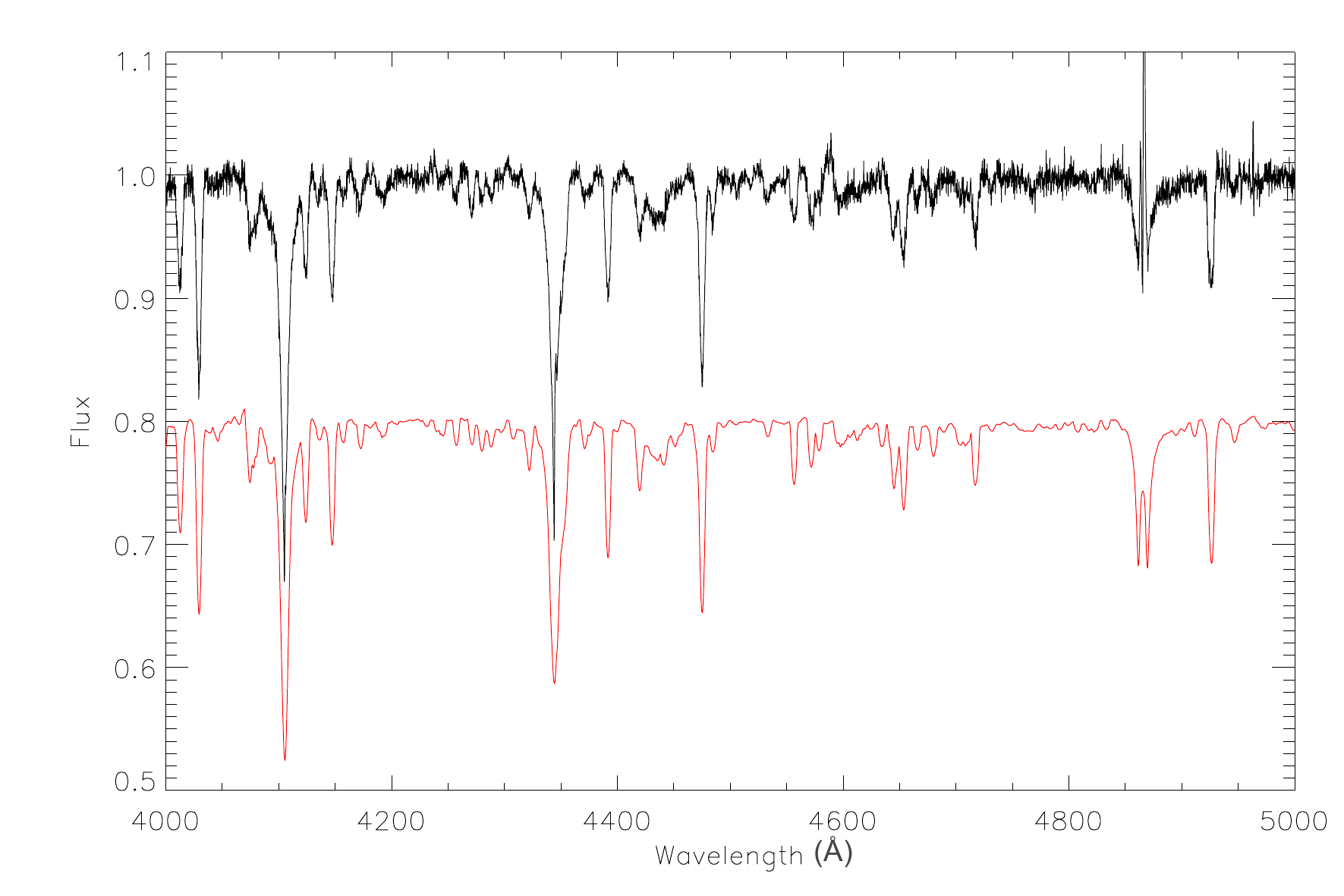,width=\linewidth,angle=0}\\
\caption{Normalised spectra for VFTS\,826 (upper) and VFTS\,845 (lower), which has been convolved with an appropriate broadening function. The broad absorption near 4430\AA \, is interstellar. }
\label{826}
\end{figure}

\subsection{ {\sc fastwind} analysis} \label{s_comp}

In order to investigate the reliability of our {\sc tlusty} analyses, we have undertaken complementary analyses using  the NLTE atmosphere/spectrum synthesis code {\sc fastwind}  (see  \citealt{san97} and \citealt{pul05} for previous versions, and  \citealt{riv12} for a summary of the latest version) and applying  a simple fit-by-eye method to determine the stellar and wind parameters.

Twelve single stars were considered, covering a range of nitrogen abundances (binary objects were not considered in order to keep the comparison as simple as possible). All models were calculated using a background metallicity, Z = 0.5\,\Zsun, appropriate to the global metallic abundances of the LMC \citep[see e.g.][and references therein]{hun07}. The adopted atmospheric parameters and nitrogen abundances in Table \ref{t_Atm} were used as starting initial estimates for the model calculations and then iterated if necessary. We note that, although {\sc fastwind} is capable of calculating microturbulent velocities, because of the method adopted in the current {\sc fastwind} analysis, it was not possible to redetermine them and therefore the {\sc tlusty} estimates were used. Additionally, the {\sc fastwind} analysis explicitly estimated the helium abundance rather than the fixed value of 0.1 (by number of atoms relative to hydrogen) adopted in the {\sc tlusty} analysis. Subsequently one target, VTFS\,533 was identified as a potential LBV candidate, (Walborn et. al. in prep.) and was therefore excluded from the analysis.

Effective temperatures were determined using the helium (\ion{He}{i} / \ion{He}{ii}) and nitrogen  (\ion{N}{ii} / \ion{N}{iii}) ionisation balances (for spectral types earlier than B2) or from the  {\ion{He}{i} 4471 line \footnote{As shown by \citet{lef07}, for early B-types this line provides a good diagnostic for gravity but at  \teff below 15~kK it becomes progressively independent of gravity and can be used to constrain the effective temperature.}. Surface gravities  were inferred  from the Stark-broadened wings of H$\gamma$ and H$\delta$, whilst the nitrogen content was constrained using the available \ion{N}{ii}  and \ion{N}{iii} lines and giving greater weight to those which are believed to be relatively  unblended \citep[see Table 2 of ][]{riv12}. The wind strength parameter $Q$ 

\citep{pul96}  was roughly constrained from the best fit to H$\beta$. A detailed investigation of the wind properties of our stellar sample was outside the scope of the present investigation and hence we did not attempt to refine these estimates. The typical uncertainties in the {\sc fastwind} analysis are comparable to those in the {\sc tlusty} analysis, being $\pm$ 1000 K in \teff,  $\pm$0.1~dex in \logg and $\pm$ 0.2~dex in nitrogen abundance.

The comparison summarised in Table \ref{t_TL_FW} is encouraging. For five targets (VFTS\,028, 069, 270, 845, 867), the two analyses led to effectively identical results. For five other targets (VFTS\,082, 165, 431, 578, 831) the agreement is satisfactory with differences $\leq$1000 K in \teff, $\leq$0.1 dex  in \logg, and $\leq$0.2 dex in the nitrogen abundance estimates. For two of these targets, modest helium enrichments were also found in the {\sc fastwind} analyses that may partially explain the differences.

For one target, VFTS\,087, the discrepancies are larger. Although the atmospheric parameter estimates show only a modest difference, the nitrogen abundance estimates differ by approximately 0.5 dex. The estimated {\sc tlusty} atmospheric parameters lie right at the edge of the grid (and indeed those from the {\sc fastwind } analysis lie outside the {\sc tlusty} grid). For the nearest {\sc tlusty}  grid point (\teff=30\,000 K, \logg = 3.25 dex) to the {\sc fastwind} estimates, the strength of \ion{N}{ii} at 3995\AA\  leads to an abundance estimate of 7.64 dex; extrapolation of the {\sc tlusty} results to the {\sc fastwind} atmospheric parameters would then imply an abundance of 7.78 dex. Both these values are in much better agreement with the {\sc fastwind} estimate of 7.80 dex. 
Hence we conclude that the discrepancy in the nitrogen abundance estimates arises from the different adopted atmospheric parameters with the results being particularly sensitive for this star which is relatively close to the Eddington limit.

In summary, for most of our sample the agreement between the two approaches is excellent. For the more extreme supergiants that lie close to the Eddington limit (see Table \ref{t_Atm}), the nitrogen abundance estimates will be less secure.
\begin{table*}

\caption{Comparison of nitrogen abundances from {\sc tlusty} and {\sc fastwind} analyses. The former are taken from Table \ref {t_Atm}, and the {\sc tlusty} microturbulences were used for the {\sc fastwind} analysis. Abundances are presented on the scale log(X/H) + 12.}\label{t_TL_FW}

\begin{center}
\begin{tabular}{lllllllllll}
\hline\hline
Star &  Spectral Type & \multicolumn{4}{c}{\sc tlusty} & \multicolumn{4}{c}{\sc fastwind}\\
&&\teff  &	 \logg&	 \vt &  N  &\teff  &	 \logg&	 He & N
 \\
\hline
028	& B0.7 Ia Nwk			& 24\,000		&2.75	&14			&7.32	
						& 24\,000		&2.75		&0.10	&7.36\\
069	& B0.7 Ib-Iab			& 23\,500		&2.75	&15			&7.93	
						& 24\,000		&2.75		&0.10	&7.93\\
082	& B0.5 Ib-Iab			& 25\,500		&3.00	&13			&7.89
						& 26\,500		&3.00		&0.15	&8.10\\
087	& O9.7 Ib-II			& 29\,000		&3.25	&12			&7.29
						& 30\,000		&3.20		&0.10	&7.80\\
165	& O9.7 Iab			& 27\,000		&3.00	&14			&7.86
						& 28\,000 		&3.00		&0.15	&8.16\\
270	& B3 Ib	    			& 17\,500		&2.90	&1			&7.57
						& 17\,500		&2.90		&0.10	&7.57\\
431	& B1.5 Ia Nstr			& 19\,000		&2.35	&18			&7.98
						& 19\,000		&2.40		&0.10	&8.10\\
578	& B1.5 Ia Nwk			& 21\,500		&2.75	&11			&7.07
						& 21\,500		&2.75		&0.10	&6.96\\
831	& B5 Ia				& 14\,000		&2.10	&21			&8.02
						& 13\,000		&1.85		&0.10	&8.22\\
845	& B1 II				& 23\,500		&3.25	&7			&7.54
						& 23\,500		&3.25		&0.10	&7.54\\
867	& B1 Ib Nwk			& 24\,500		&3.15	&11			&7.16
						& 26\,000		&3.25		&0.08	&7.16\\
\hline
\end{tabular}
\end{center}

\end{table*}

\section{Stellar parameters} \label{s_st_par}

Other stellar parameters can be estimated from the observational data and atmospheric parameters; these include the luminosity (L) and mass, the latter often being referred to as the spectroscopic mass, \Msp. Additional parameters can be inferred from comparison with evolutionary tracks including age ($t$) and evolutionary mass, \Mev. These estimates are summarised in Table \ref{t_obs} and discussed below.

\subsection{Observationally inferred parameters: Luminosities and spectroscopic masses} \label{Lum_SM}

When available, near-infrared ($J$- and $K_{\rm s}$-band) photometry from VISTA \citep{cio11} and 2MASS \citep{cut03} was used to estimate stellar luminosities.  This was supplemented by the visual $B,V$ magnitudes compiled in \cite{eva11}. For 2MASS photometry, only stars with $J,K_{\rm s}$ magnitudes having S/N $>$\,7 and a mean error estimate of $<$\,0.15 mag were considered. Model colours ($J-K_{\rm s}$  and $B-V$) and bolometric corrections for the appropriate filters were computed using  the theoretical spectra of \citet{cas03} by adopting the nearest available model in the  \teff, \logg\ grid for each individual object (see Table \ref{t_Atm}), with log\,[M/H]\,=\,$-$0.5. The reddening follows from the difference between the observed and model colours, whilst extinction in the $V$ and $K_{\rm s}$\ bands was estimated using a reddening law with $R_{V}$\,=\,3.5 \citep[consistent with that adopted by][]{dor13}, following the  parametrisation of \citet{car89}. \footnote{In the course of this work a new set of extinction laws have become available from \citet{mai14}; in the context of this investigation, use of these laws did not result in significant ($\ge$\,0.01\,dex) changes in the calculated  luminosities.} The stellar luminosity was then calculated using a distance modulus $\mu$\,=\,18.5\,$\pm$\,0.1, as adopted in \citet{eva11}. The infrared extinctions ($A_k $) for each star are listed in Table \ref{t_obs}. For the VISTA estimates the  formal errors ranged from 0.01 to 0.09 but in most cases were $\approx$ 0.02. For the 2MASS $A_k$ values, the formal errors ranged from 0.02 to 0.12, averaging $\approx$ 0.05.

The luminosities (summarised in Table \ref{t_obs}) deduced from the infrared photometry are in good agreement with a mean difference (Vista-2Mass) of $-$0.01$\pm$0.06 dex. By contrast, the visible luminosities are systematically lower with a mean difference of -0.10$\pm$0.12 dex (VISTA) and  -0.10$\pm$0.11 dex  (2MASS). This may reflect the greater sensitivity of the visible estimates to the adopted global value of $R_{V}$ \citep[see][for a detailed investigation of the magnitude and variation of this parameter in 30 Doradus region]{mai14}. 

Hence, where available, we have normally adopted the IR luminosities (with preference given to the VISTA photometry). However for three targets, VFTS\,420, 590, and 591, inspection of HST imaging \citep[see][for details]{eva11} indicated that the 2MASS photometry might be affected by close companions and the visible luminosities were therefore adopted. Including those targets where no IR photometry was available (VFTS\,826, 831), visible luminosities were therefore adopted for five of our targets. 

From the estimated luminosity, effective temperature, and gravity, it is possible to infer a spectroscopic mass, \Msp, as summarised in Table \ref{t_obs}. These will be subject to considerable uncertainty with for example, random uncertainties of 0.1 dex in the luminosity, 5\% in the effective temperature and 0.1 dex in logarithmic gravity translating into uncertainties of 13\%,  22\%, and 26\%, respectively. Hence a typical {\em random} uncertainty of 40\% could be considered appropriate, although there may be additional uncertainties due to systematic errors in the estimation of the luminosities or atmospheric parameters.

\begin{table*}

\caption{Estimates of the stellar luminosity ($L$) and spectroscopic mass (\Msp) for our sample. The infrared extinctions ($A_k$) used to calculate these values are included.
Also listed are the stellar parameters obtained from single-star evolutionary models. Stellar ages ($t$) are in years, whilst both the initial (\Mor) and current (\Mev) evolutionary masses are estimated.}\label{t_obs}

\begin{center}
\begin{tabular}{llcccccccccc}
\hline\hline
Star   &  ST & \multicolumn{2}{c}{$A_k$} & \multicolumn{4}{c}{~~~~~~~$\log$\ L/\Lsun} &   \Msp/\Msun &    \Mev/\Msun & \Mor/\Msun &	 $\log t$ 
 \\
         &      & VISTA & 2MASS  & Visible  & VISTA & 2MASS & Adopted\\

\hline
003	& B1 Ia$^+$			&-&0.24 				&  -    	&   -        	 & 6.03	& 6.03	& 71		&	44	&	49	&	6.56			\\
028	& B0.7 Ia Nwk	&	-			&	0.33		& 5.51	&   -		& 5.76	& 5.76 	& 40		&	44	&	48	&	6.56 		\\
060	& B1.5 II-Ib((n))	 &	0.42	&	0.37				& 4.75	& 4.82	& 4.85	& 4.82	&  8		&	23	&	23	&	6.84			\\
069	& B0.7 Ib-Iab	&	0.29	&	0.29				& 5.34	& 5.59     	& 5.61	& 5.59 	& 29		&	37	&	40	&	6.62	  		\\
082	& B0.5 Ib-Iab	&	0.15	&	0.18				& 5.09	& 5.26	& 5.32	& 5.26 	& 17		&	30	&	31	&	6.71 		\\
087	& O9.7 Ib-II	&	0.09	&	0.09				& 5.16	& 5.27	& 5.29	& 5.27 	& 19		&	30	&	31	&	6.68 		\\
165	& O9.7 Iab	&	0.20	&	-				& 5.33	& 5.46     	& -		& 5.46 	& 22		&	39	&	42	&	6.59 		\\
178	& O9.7 Iab	&	0.14	&	0.12				& 5.47	& 5.60	& 5.59	& 5.60 	& 33		&	35	&	37	&	6.63 		\\
232	& B3 Ia	&0.30 	&0.32 					& 4.93	& 4.90	& 4.93	& 4.90	& 11		&	21	&	21	&	6.88 		\\
261	& B5 Ia       &	-			&	0.16			& 5.02	&  -		& 5.14	& 5.14	& 21		&	20	&	20	&	6.91			\\
269	& B8 Ia      &	-			&	0.07			& 4.62	&  -		& 4.71	& 4.71	&  8		&	19	&	19	&	6.93			\\
270 	& B3 Ib	 &	0.06	&	-			   		& 4.36	& 4.51	& -		& 4.51	& 11		&	12	&	13	&	7.18			\\
293	& B2 III-II(n)e		&	0.30		&	-		& 4.32	& 4.84	& -		& 4.84	& 18		&	15	&	15	&	7.08			\\
302	& B1.5 Ib  		&	0.31	&	0.29				& 4.62	& 4.78	& 4.78	& 4.78	& 11		&	18	&	18	&	6.96			\\
307	& B1 II-Ib		&	0.11	&	0.15				& 4.96	& 5.09	& 5.11	& 5.09	& 13		&	26	&	27	&	6.79			\\
315	& B1 Ib	&	0.11		&	0.16				& 4.63 	& 4.64	& 4.69	& 4.64	&  9		&	17	&	17	&	6.98			\\
417	& B2 Ib	& 0.35 &0.49 						& 4.51	& 4.73	& -	 	& 4.51	&  4		&	23	&	24	&	6.83			\\
431	& B1.5 Ia Nstr	&	-		&	0.23			& 5.60	&  -		& 5.79	& 5.79	& 43		&	36	&	40	&	6.64			\\
458	& B5 Ia$^+$p	&	-		&	0.38			& 5.59	&  -		& 5.64	& 5.64	& 25		&	47	&	54	&	6.54			\\
533	& B1.5 Ia$^+$p Nwk		&	-		&0.20	 & 5.89   	&  -		& 5.88	& 5.88	& 37		&	-	&	$\!\!\!\!>$60	&  - 		\\
578	& B1.5 Ia Nwk		578	&	0.36		&	0.39	& 5.19	& 5.35	& 5.40	& 5.35	& 24		&	26	&	27	&	6.78			\\
590	& B0.7 Iab	-	&	-			&	0.25	& 5.87	&  -		& 6.01	& 5.87	& 57		&	36	&	39	&	6.63			\\
591	& B0.2 Ia		-	&	-			&	0.27	& 5.91	&  -		& 6.14	& 5.91	& 53		&	48	&	53	&	6.52			\\
672	& B0.7 II Nwk?		672	&	0.30	&	0.25		& 5.09	& 5.23	& 5.22	& 5.23	& 18		&	27	&	29	&	6.75			\\
696	& B0.7 Ib-Iab Nwk	-	&	-		&	0.20	& 5.42	&  -		& 5.64	& 5.64	& 33		&	37	&	40	&	6.62			\\
714	& B1 Ia: Nwk	&	0.22	&	-				& 4.77	& 4.74	& -		& 4.74	&  7		&	22	&	23	&	6.84			\\
732	& B1.5 Iap Nwk		&	-		&	0.28		& 5.41	&  -		& 5.61	& 5.61	& 33		&	32	&	35	&	6.69			\\
745	& B2.5 II-Ib	&	0.32	&	-				& 3.99	& 4.20	& -		& 4.20	&  4		&	15	&	15	&	7.07			\\
826  & B1 IIn	&- &- 							& 4.85  	&  -		& -		& 4.85	& 17		&	16	&	16	&	7.04			\\
831 	& B5 Ia	&- &- 							& 5.10	&  -		& -		& 5.10	& 17		&	21	&	22	&	6.87			\\
841	& B2.5 Ia & 0.46 &0.51 						& 5.09	& 5.10	& 5.14	& 5.10	& 14		&	25	&	26	&	6.80			\\
845	& B1 II	&	0.23	&	0.21					& 4.64	& 4.78	& 4.79	& 4.78	& 14		&	16	&	16	&	7.04			\\
855 	& B3 Ib	&	0.16	&	0.05					& 4.40	& 4.32	& 4.20	& 4.32	&  6		&	14	&	14	&	7.11			\\
867	& B1 Ib Nwk	&	0.18	&	0.11				& 4.91 	& 4.93	& 4.88	& 4.93	& 14		&	20	&	21	&	6.90			\\
\hline
027	& B1 III-II		&	0.14	&	0.01				& 4.63	&  4.63	& 4.46	& 4.63	& 12		&	14	&	14	&	7.12			\\
045	& O9.7 Ib-II Nwk	&	0.34	&	0.39			& 5.20	&  5.24	& 5.28	& 5.24	& 18		&	30	&	31	&	6.68 		\\
291 & B5 II-Ib	&0.26  &0.18 						& 4.09	&  4.31	& 4.23	& 4.31	&  6		&	14	&	14	&	7.11			\\
420	& B0.5 Ia Nwk		&	-	&	0.22			& 5.84 	&  -		& 5.95	& 5.84	& 57		&	35	&	37	&	6.63			\\
423	& B1 Ia: Nwk	&	-			&	0.40		& 5.59     	&  -		& 5.72	& 5.72	& 38		&	36	&	39	&	6.64			\\
430 & B0.5 Ia+((n)) Nwk	&	0.50	&	0.45			& 5.44	& 5.56	& 5.59	& 5.56	& 18		&	-	&	$\!\!\!\!>$60	& - 		\\
450	& O9.7 III: + O7::		&	0.25	&	0.25		& 5.56 	& 5.67	& 5.67	& 5.67	& 36		&	39	&	42	&	6.59			\\
525	& B0 Ia	&	0.21		&	-				& 5.44	& 5.48	& -		& 5.48	& 23		&	39	&	42	&	6.59			\\
541	& B0.5 Ia Nwk	&	0.19	&	0.23				& 5.54	& 5.57	& 5.63	& 5.57	& 31		&	34	&	36	&	6.65			\\
576	& B1 Ia Nwk	&	0.51	&	0.60				& 5.18	& 5.31	& 5.39	& 5.31	& 16		&	32	&	35	&	6.69			\\
652	& B2 Ip + O9 III: 	&	0.26	&	0.23			& 5.14	& 5.16	& 5.19	& 5.16	& 10		&	44	&	49	&	6.56			\\
675	& B1 Iab Nwk		&	0.25	&	0.30			& 5.09	& 5.16	& 5.20	& 5.16	& 13		&	30	&	32	&	6.72			\\
687	& B1.5 Ib((n)) Nwk	&	0.20	&	0.23			& 5.04	& 4.95	& 4.97	& 4.95	& 10		&	25	&	25	&	6.80			\\
733	& O9.7p			&	0.21	&	0.12			& 5.26	& 5.34	& 5.23	& 5.34	& 21		&	32	&	34	&	6.65			\\
764	& O9.7 Ia Nstr		&	-		&	0.13		& 5.96	&  -		& 5.85	& 5.85	& 50		&	44	&	48	&	6.54			\\
779	& B1 II-Ib	&	0.22		&	0.36				& 4.62	& 4.73	& 4.86	& 4.73	& 11		&	17	&	17	&	7.00			\\
827	& B1.5 Ib		&	0.41&	-				& 4.71	& 5.03	& -		& 5.03	& 28		&	15	&	15	&	7.08			\\
829	& B1.5-2 II	&	0.21&	-				& 4.91	& 4.78	& -		& 4.78	& 11		&	18	&	18	&	6.96			\\
\hline
\end{tabular}
\end{center}
\end{table*}

\subsection{Evolutionary inferred parameters: Ages and Masses}\label{s_ev_est}

Comparison of our adopted atmospheric parameters with those predicted from grids of evolutionary models allow additional stellar parameters to be estimated, specifically the age ($t$) from the zero age main sequence (ZAMS), the current mass, \Mev\  and the ZAMS mass, \Mor. The difference between \Mev\ and \Mor\ is strongly controlled by the mass-loss recipe used in the evolutionary calculations, but since the difference is low in most cases, this plays very little role in the present sample. We have used the  grid of models of \citet{bro11a} as these include a metallicity appropriate to the LMC, and cover a wide range of initial equatorial velocities ($0\le$\vor$\la 600$\kms) and masses ($0\le$\Mor $\le 60$\Msun); they have also been used in other papers in this series, thereby ensuring consistency. We have chosen to use those models with \vor$\simeq 225$\ \kms as this was compatible with the mean equatorial velocity inferred for  (near) main sequence stars observed in the Tarantula survey \citep{duf12, ram13}. The uncertainties arising from this choice will be discussed below.

Ages, masses, and gravities were extracted from the appropriate models for effective temperatures ranging from 11\,000 to 35\,000 K (at intervals of 500~K, corresponding to the rounding of our \teff estimates). The grid of \cite{bro11a} had a sufficiently fine mesh that errors from the (linear) interpolation should be negligible. Ages and masses were then estimated from a quadratic interpolation in logarithmic gravity. Use of either linear or spline interpolation led to very similar results, implying that the choice of interpolation methodology is unlikely to be a significant source of error. Two targets (VFTS\,430 and 533) lay outside this grid implying that their initial mass was greater than 60\Msun, (see Fig. \ref{f_HRD} lower plot). The BONNSAI facility \citep{sch14} was used to try to estimate parameters for these stars\footnote{The BONNSAI web-service is available at www.astro.uni-bonn.de/stars/bonnsai}. Adopting an initial rotational velocity, \vor$\simeq 225$\, no solution could be found; although allowing this parameter to vary would have led to solutions, this was not pursued to ensure compatibility with the other estimates. Some differences in evolutionary mass estimates can be seen in Fig \ref{f_HRD} between the two plots, one showing effective temperature against luminosity and the other effective temperature against $\log g$. We have adopted our evolutionary masses from the effective temperature verses $\log g$ plot. This discrepancy has also been discussed in \cite{lan14} and has been attributed to the stars becoming over-luminous as a result of, for example, binary evolution or rapid rotation. 

We have investigated the effects of errors in the atmospheric parameters by considering errors of $\pm$1000 K in \teff\ and $\pm$0.1 dex in \logg. These are representative of the {\em random} errors in the fitting procedures and there may be additional systematic errors due to, for example, inadequacies in the physical assumptions made. Additionally, increasing the effective temperature by 1000 K would generally lead to an increase in \logg of 0.1 dex, but here we have simply varied one variable at a time. Five representative single supergiants were considered covering a range of luminosity classes (from Ib to Ia$^+$) and spectral types from B0.7 to B5, with the results summarised in Table \ref{t_ev_err}. Implied percentage uncertainties in the masses range from 20-40\% and increase as the Eddington limit is approached. For lifetimes, there is less variation between the targets with uncertainties being somewhat smaller.

\begin{table}

\caption{Effects of the estimated {\em random} errors in the atmospheric parameters on the estimation of current masses and lifetimes from evolutionary models. In each case the relevant atmospheric parameter has been increased although tests showed that decreases led to similar results but with the sign changed. All mass estimates are in units of the solar mass. For convenience the evolutionary masses have been reproduced from Table \ref{t_obs}.}\label{t_ev_err}

\begin{center}
\begin{tabular}{lllcccccc}
\hline\hline
Star & ST & &\multicolumn{2}{c}{$\Delta$\teff = $+$1000 K}  & \multicolumn{2}{c}{$\Delta$\logg = $+$0.1 dex}     
 \\	
&& \Mev & $\Delta$\Mev & $\Delta\log$ t & $\Delta$\Mev & $\Delta\log$ t 
\\ \hline
270	& B3 Ib   		& 15		& +2 & $-$0.07 & $-$1 & +0.08    \\
261	& B5 Ia    		& 20		& +4 & $-$0.09 & $-$3 & +0.08    \\
578	& B1.5 Ia   	& 26		& +4 & $-$0.06 & $-$4 & +0.08    \\
590	& B0.7 Iab	& 36		& +12 & $-$0.11 & $-$5 &  +0.07  \\
003	& B1 Ia$^+$		& 44		& +13 & $-$0.08 & $-$8 &  +0.08  \\
\hline
\end{tabular}
\end{center}
\end{table}
We have also investigated the sensitivity of our estimates to the initial rotational velocity and to the adopted grid. For the former, we considered models with initial rotational velocities of 0 and 400 \kms from the grid of \cite{bro11a}. For the latter, we used the database of models provided by the Geneva stellar evolution group \footnote{see, http://obswww.unige.ch/Recherche/evol/-Database- for the Geneva stellar evolutionary models}. This provides a wide range of models but regrettably not at the metallicity (Z$\simeq$0.006) and range of stellar masses appropriate to our sample. Grids are available at metallicities, Z, of 0.014 and 0.002 and we have used both to gauge the sensitivity of the estimates to the adopted metallicity. Additionally, these models are only available at two initial rotation velocities of zero and forty per cent of the critical velocity (the latter corresponding to a typical value of \vor$\simeq$260 \kms). The spacing of the grid points is also relatively coarse (with \Mor\ being 15, 20, 25, 32, 40, 60\Msun) and hence we have only undertaken a linear interpolation to estimate parameters.

We have considered the same five  single supergiants as in Table \ref{t_ev_err} and summarise the results in Table \ref{t_ev_comp}. Generally the agreement is satisfactory between all models when the target lies well away from the Eddington limit (i.e. low effective temperature and/or high gravity; VFTS\,270, 261, and  578). For targets nearer the Eddington limit (see Table \ref{t_Atm}), the choice of the original rotational velocity and adopted grid leads to significant variations in both masses and lifetimes. 

In summary, evolutionary mass estimates suffer significant uncertainties from possible errors in the atmospheric parameter determinations. Additionally, it is important that models appropriate to the metallicity regime of the targets are used. For stars near the Eddington limit, the initial rotational velocity becomes critical. We note that \citet{duf12} and \cite{ram13}  predicted targets with a wide range of rotational velocities including \vor$\ga$ 400 \kms for their B-type and O-type non-supergiant VFTS samples, respectively. As the original rotational velocities of our supergiants are currently unknown, the implied evolutionary masses must be considered with considerable caution. {\em This caveat also applies to all previous studies of early-type supergiants where the initial stellar rotational velocity has not been explicitly considered}.

Finally it should be noted that all the above analyses make the crucial assumption that single-star evolutionary models are appropriate and that there are no significant  effects due to binarity \citep[see, for example,][]{deM11, deM13, deM14}.

\begin{table}

\caption{Estimates of the stellar parameters obtained from different single-star evolutionary grids. B11 is from the grid of \cite{bro11a} for an LMC metallicity. E12Z014 is from \cite{eks12} for Galactic metallicity, with E12Z002 being for a lower (near SMC) metallicity. The rotating models of \cite{eks12} are for an intital rotational velocity of 40\% of the critical value that translates to \vor$\sim$260 \kms. Masses are in units of the solar mass and time is in years.}\label{t_ev_comp}

\begin{center}
\begin{tabular}{lllcccccc}
\hline\hline
Star & ST & Grid  & \vor &\Mev & \Mor & $\log t$     
 \\	\hline

270	& B3 1b        	& B11 		& 0     	& 12    	& 12		& 7.21	\\
	&		     	& B11		& 225	& 12		& 13		& 7.18	\\
	&        		& B11 		& 400 	& 12		& 12		& 7.23	\\
	&			& E12Z002	& 0		& 12        	& 13        	& 7.14      \\
	&			& E12Z014	& 0   	& 14		& 14		& 7.09	\\
	&			& E12Z002	& 260	& 12		& 12		& 7.28	\\
	&			& E12Z014	& 260	& 12		& 12		& 7.24	\\
\hline
261	& B5 Ia       	& B11 		& 0     	& 20    	& 20		& 6.91	\\
	&			& B11		& 225	& 20		& 20		& 6.91	\\
        &			& B11 		& 400 	& 19		& 20		& 6.94	\\
	&			& E12Z002	& 0		& 18        	& 18       	& 6.99     \\
	&			& E12Z014	& 0   	& 19		& 20		& 6.90	\\
	&			& E12Z002	& 260	& 15		& 15		& 7.16	\\
	&			& E12Z014	& 260	& 18		& 18		& 7.03	\\

\hline
578	& B1.5 Ia 		& B11 		& 0     	& 26    	& 27		& 6.79	\\
	&			& B11		& 225	& 26		& 27		& 6.78	\\
	&			& B11 		& 400 	& 24		& 25		& 6.86	\\
	&			& E12Z002	& 0		& 26		& 26		& 6.80 	\\
	&			& E12Z014	& 0		& 28		& 29		& 6.75 	\\
	&			& E12Z002	& 260	& 24		& 24		& 6.90	\\
	&			& E12Z014	& 260	& 23		& 24		& 6.92	\\
\hline
590	& B0.7 Iab	& B11 		& 0     	& 39    	& 42		& 6.60	\\
	&			& B11		& 225	& 36		& 39		& 6.63	\\
        &			& B11 		& 400 	& 30		& 33		& 6.76	\\
	&			& E12Z002	& 0		& 35		& 36		& 6.69	\\
	&			& E12Z014	& 0		& 36 	& 40		& 6.65 	\\
	&			& E12Z002	& 260	& 30		& 31		& 6.82	\\
	&			& E12Z014	& 260	& 26		& 29		& 6.86	\\
\hline	
003	& B1 Ia$^+$		& B11  		& 0     	& 46    	& 50		& 6.55	\\
	& 			& B11  		& 225 	& 44		& 49		& 6.56	\\
        &			& B11 		& 400 	& 30		& 33		& 6.76	\\
	&			& E12Z002	& 0		& 38		& 39		& 6.67	\\
	&			& E12Z014	& 0		& \,\,50:	& \,\,57:	& 6.55	\\
	&			& E12Z002	& 260	& 35		& 36		& 6.77	\\
	&			& E12Z014	& 260	& 28		& 32		& 6.83	\\
\hline
\end{tabular}
\end{center}
\end{table}

\section{Discussion}\label{s_discussion}

\subsection{Hertzsprung-Russell diagram}\label{ss_HRD}

In Fig. \ref{f_HRD}, the effective temperature and luminosity estimates from Tables \ref{t_Atm} and \ref{t_obs} have been used to construct a Hertzsprung-Russell diagram. Also shown are the evolutionary tracks of \cite{bro11a} for an initial rotational velocity of approximately  \vor$\simeq 225$\ \kms consistent with the value adopted in Section \ref{s_ev_est}. To illustrate how the evolutionary masses listed in Table \ref{t_obs} were estimated an analogous plot is shown (lower plot, Fig. \ref{f_HRD}), with effective temperature against logarithmic surface gravity. 

In Fig. \ref{f_HRD_zoom}, we show a zoomed-in H--R diagram, with a thick TAMS (terminal age main sequence) line plotted. Surprisingly the distribution of binary stars extends beyond the TAMS. As these are likely pre-interaction systems this strongly suggests that the redwards extent of the TAMS is underestimated in our models, at least in the mass range of approximately 10-30 \Msun. In other words those binaries and single stars lying a little cooler that the TAMS may in fact be core hydrogen burning stars. Those, predominantly single, cooler stars (with \teff $\le$ 20\,000 K; highlighted in green in Fig. \ref{f_HRD_zoom}) might be interpreted as core He burning objects. The lack of binary detections would then be consistent with a scenario where they are blue-loop, or post red supergiant (RSG), stars since post-TAMS evolution (during the RSG phase) removes practically all detectable binaries beyond the TAMS \citep[][ in particular their Figure 2]{deM14}. There is one apparent exception in our sample, namely VFTS\,291, which may be worth analysing in detail.

If it is true that the cooler stars in the sample are post-RSG objects then, since their evolutionary masses are derived without making this assumption (Fig.\ref{f_HRD}), their spectroscopic masses should be significantly lower \citep[see][]{len05}.The evolutionary and spectroscopic mass estimates (taken from Table \ref{t_obs}) are shown in Fig. \ref{f_Ev_Spec}. For most of our sample, the evolutionary mass estimates are larger, with only twelve targets having higher spectroscopic mass estimates. This is similar 
to results found by other authors \citep[e.g.][]{her02, tru04, tru05, sea08} from studies of supergiants in the Galaxy and SMC, using non-rotating \citep{sch92}, and rotating \citep{mae01} evolutionary models. 

Perhaps more physical insight is provided by Fig. \ref{f_mass_dis} where we plot mass discrepancy \citep{her92} against luminosity. In this figure we can think of the area between the two dashed blue lines as being the "allowed region". The lower line ($M_e = M_{s}$) exists because stars cannot be under-luminous, but only over-luminous. The upper blue dashed line is plotted to represent the mass-luminosity relation of helium stars: a star of a given mass  (as measured by $M_{s}$) cannot be more luminous than this. Reassuringly almost all the stars are contained in this allowed region. If we ignore the few outliers (see below), we can also see that the distribution becomes narrower for higher luminosity, which is in line with the model predictions (and is consistent with the observations of \cite{tru05} for supergiants in the SMC). Regarding the outliers, there are four stars above the general distribution. Those are objects where the evolutionary mass is more than three times the spectroscopic mass. Two of them are identified as binaries and it may be that their luminosities and spectroscopic masses are unreliable. The under-luminous outliers are harder to understand. Eight of these lie very close to the lower limiting line, and within their errors could easily fall back within the allowed distribution. However, four outliers remain. Except for one identified binary (at log L/\Lsun$=$5), the other three are very close to the Eddington limit (near log L/\Lsun=6), where the atmospheric parameter estimates may be unreliable.

However the present data provide, for the first time, a well-sampled view of mass-discrepancies at lower luminosities for B-type supergiants. Figure \ref{f_mass_dis}  is perhaps confusing given its mix of binaries and single stars. In addition, given the discussion of a possible extended main sequence above, we now restrict our sample to single stars, separated into stars with Teff > 20\,000 K and Teff < 20\,000 K,
this temperature being approximately the boundary of the extended TAMS (Fig.\ref{f_HRD_zoom}). The justification for excluding binaries is that we wish to minimise possible scatter due to the potential influence of secondary stars on the photometry and spectroscopic masses of that sample. The single-star mass-discrepancy diagram is shown in Fig. \ref{f_mass_dis_singles} in which we highlight these two groups of possible
pre-TAMS and post-TAMS stars. While there is still considerable scatter in the mass discrepancies, it is interesting to note that on average, mass discrepancies are larger in the potential pre-TAMS group (larger in the sense that $M_e > M_s$). 

Clearly, the evidence from the mass-discrepancies on their own is not conclusive, however given the strong supporting evidence from the distribution of binaries in the H--R diagram (Fig.\ref{f_HRD_zoom}), a reasonable hypothesis is that the main-sequence band extends at 
least until an effective temperature of  20\,000 K. While this is cooler than adopted by \cite{bro11a}, we note that the \vsini\ distribution they used to define the TAMS \citep{hun08a}, in particular the high \vsini\  stars, is also extended by the present work to approximately 20\,000 K (cf section 5.11, although an alternative interesting explanation for the drop in \vsini\ might still be bi-stability braking). Further circumstantial evidence for an extended main sequence is provided by the discovery of the dichotomy of nitrogen abundance distributions for stars on either side of this Teff, i.e. below 20\,000 K there is an absence of stars with 
normal (for the LMC) nitrogen abundance (Fig.\ref{N_atm}), all are highly enriched. 

Taken together these four pieces of evidence, namely the distribution of binaries, mass discrepancies, \vsini\ distribution, and nitrogen abundances, hint at a main-sequence width which is even broader than that adopted by \cite{bro11a}, and extending to an effective temperature of around
20\,000K at intermediate masses. We note that \cite{cas14} recently came to a similar conclusion for Galactic B-type supergiants,
  based on their distribution in the spectroscopic Hertzsprung-Russell diagram.

 \begin{figure}
\epsfig{file=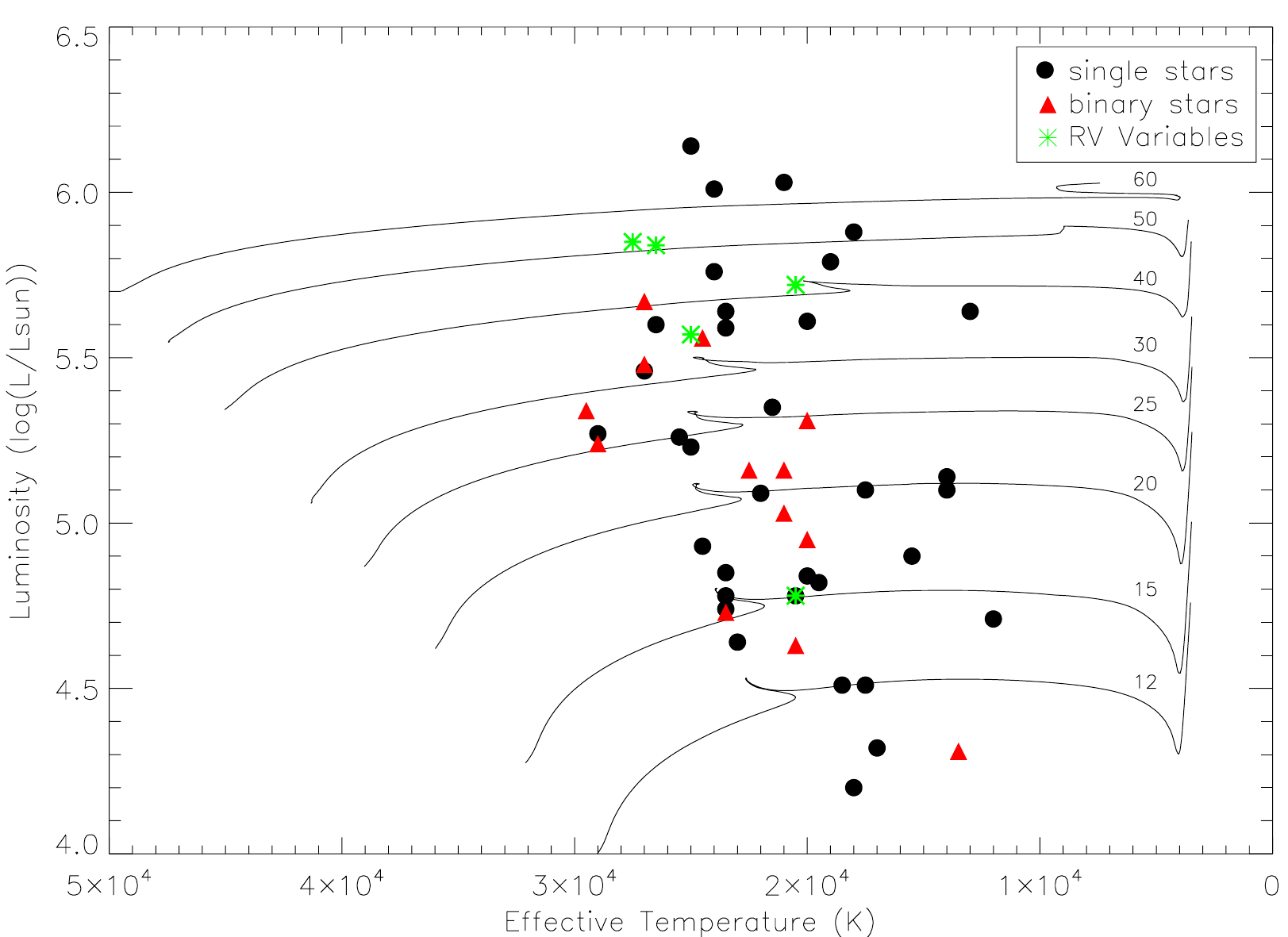,width=\linewidth,angle=0}\\
\epsfig{file=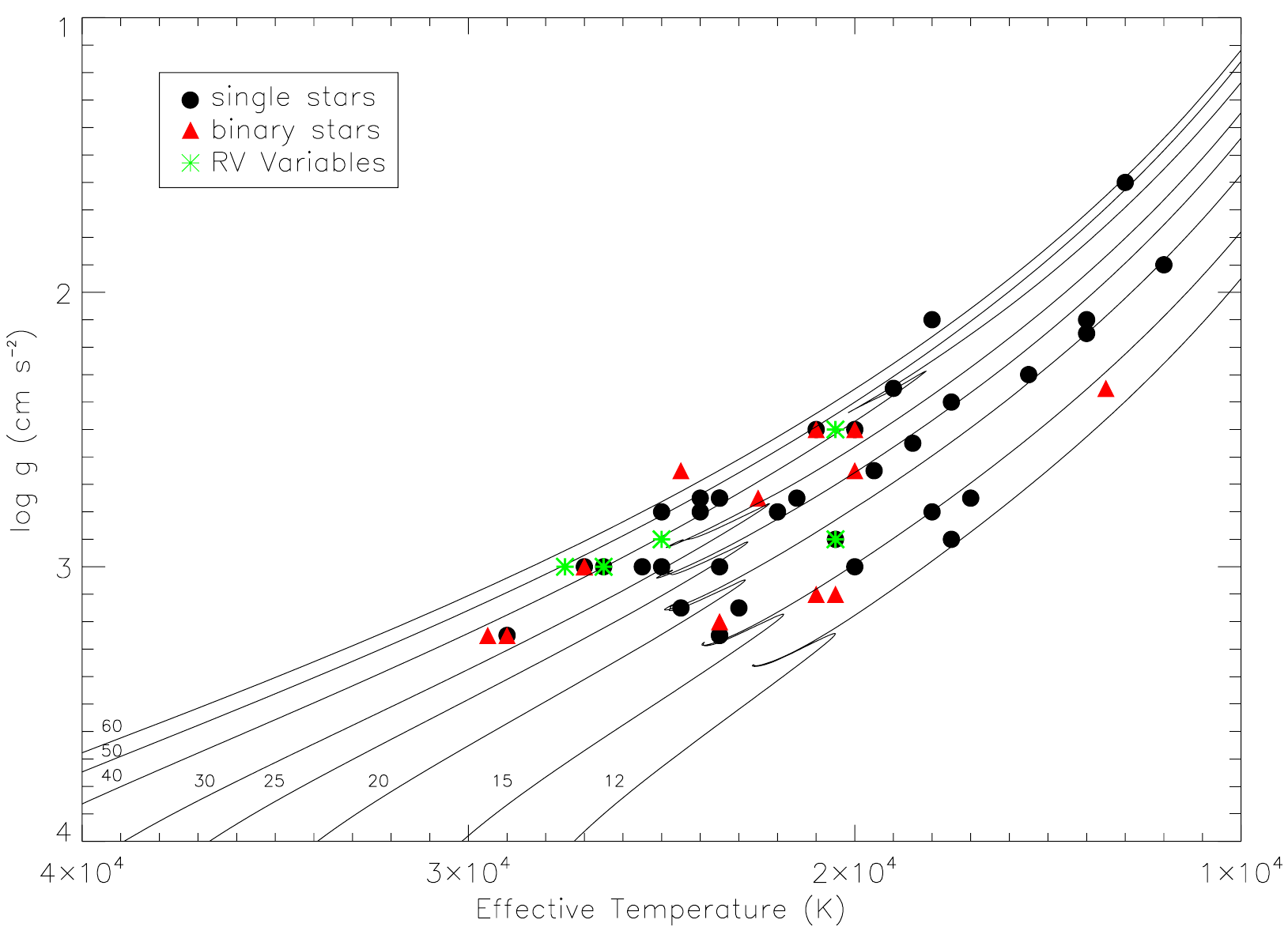,width=\linewidth,angle=0}\\
\caption{Top: classic Hertzsprung-Russell diagram for the single (circle) and binary (triangle) supergiants. Also shown are the single-star LMC evolutionary tracks of \cite{bro11a} for \vor$\simeq 225$\ \kms, with the initial mass (in units of the solar mass) given on the right hand side. Bottom: analogous plot showing effective temperatures versus logarithmic surface gravity. The same evolutionary tracks are plotted, this time with the initial masses given at the bottom of each track. The green stars in both represent the five stars referred to in Section \ref{s_binarity} as radial velocity variables. }
\label{f_HRD}
\end{figure} 

 \begin{figure}
\epsfig{file=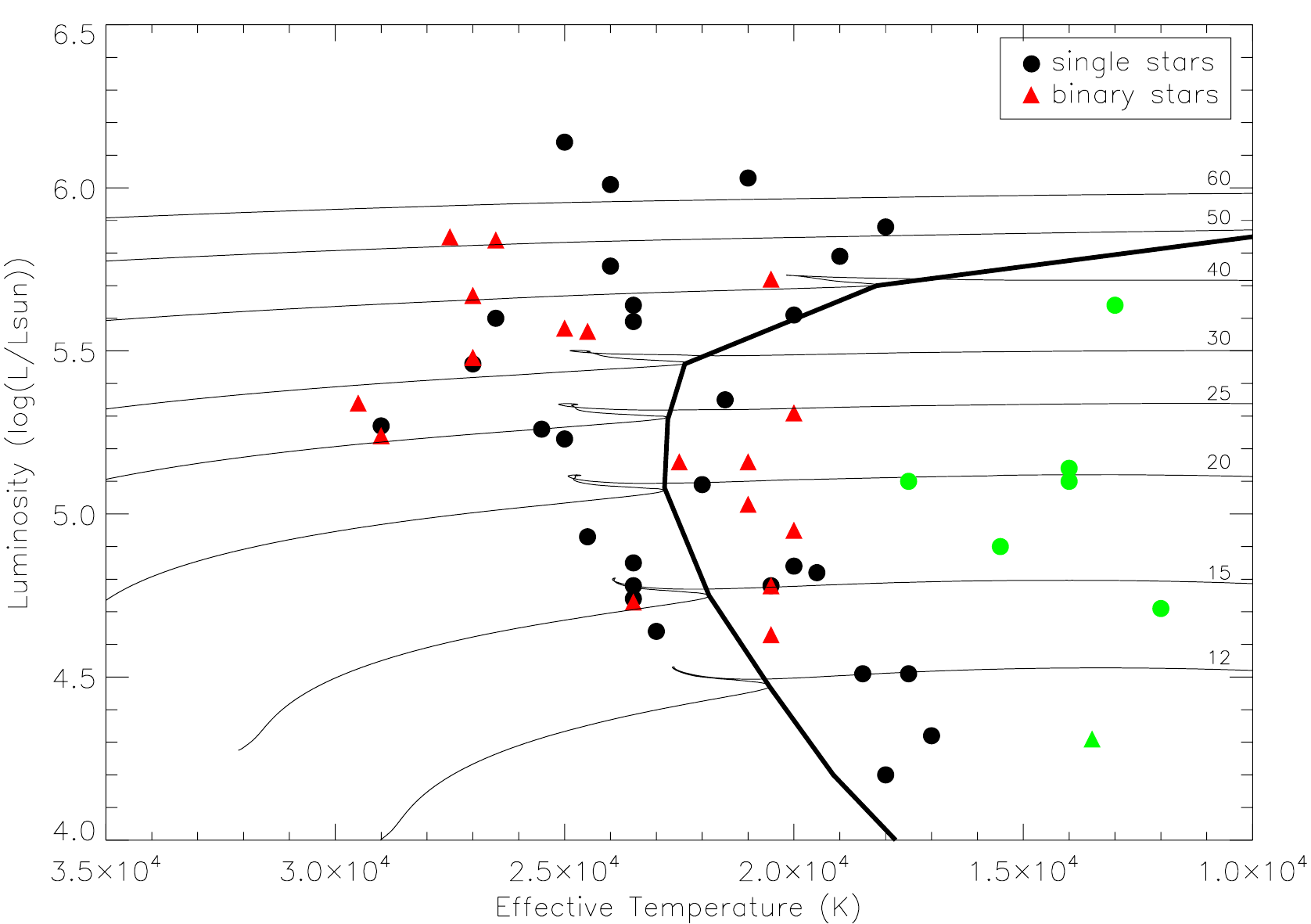,width=\linewidth,angle=0}\\
\caption{A zoomed-in version of the Hertzsprung-Russell diagram for the single (circle) and binary (triangle) supergiants. Also shown are the LMC evolutionary tracks of \cite{bro11a} for \vor$\simeq 225$\ \kms, with the initial mass (in units of the solar mass) given on the right hand side. The dark line represents the TAMS (terminal age main sequence). The stars highlighted in green are far enough from the TAMS line that they may be interpreted as core Helium burning objects. }
\label{f_HRD_zoom}
\end{figure} 

\subsection{Binary fraction}

For our supergiant sample (see Table \ref{t_Targets}), approximately one third (18 out of 52 targets) have been classified as binaries including two double-lined binaries. This fraction is consistent with the (0.35$\pm$0.03) binary fraction found by \citet{san13} for the O-type stars in 30 Doradus, which would normally be the precursors of B-type supergiants. However, the two analyses used different criteria and, in particular, several of our supergiants were classified as binaries although they would have fallen below the threshold for the O-type binary criteria. We note that \citet{san13} estimated a true binary fraction of 0.51$\pm$0.04 after allowing for incompleteness. 

 \begin{figure}
\epsfig{file=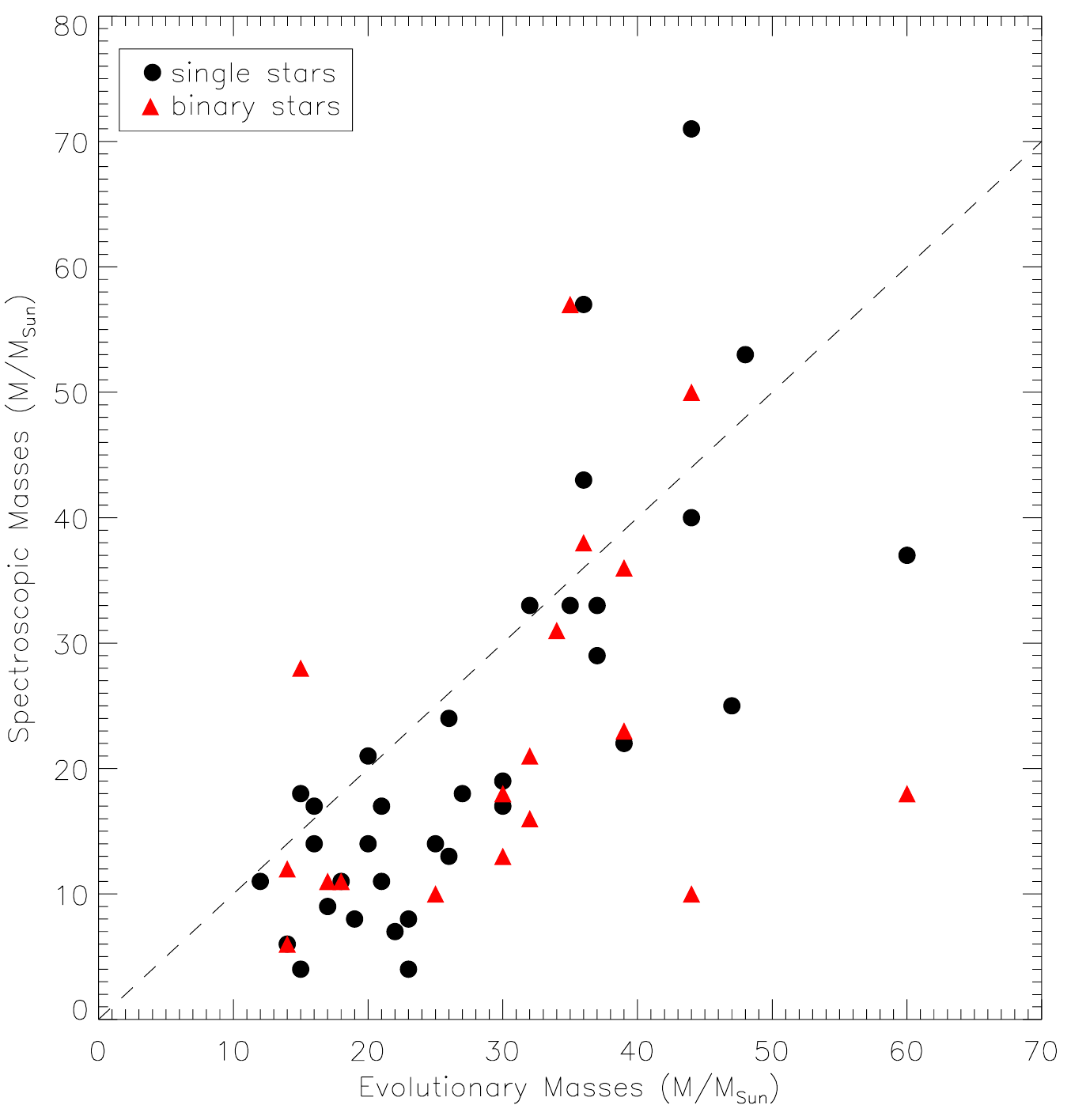,width=\linewidth,angle=0}\\
\caption{Comparison of evolutionary and spectroscopic stellar masses for our sample of B-type Supergiants. The dashed line indicates a 1:1 correspondance.}
\label{f_Ev_Spec}
\end{figure} 

 \begin{figure}
\epsfig{file=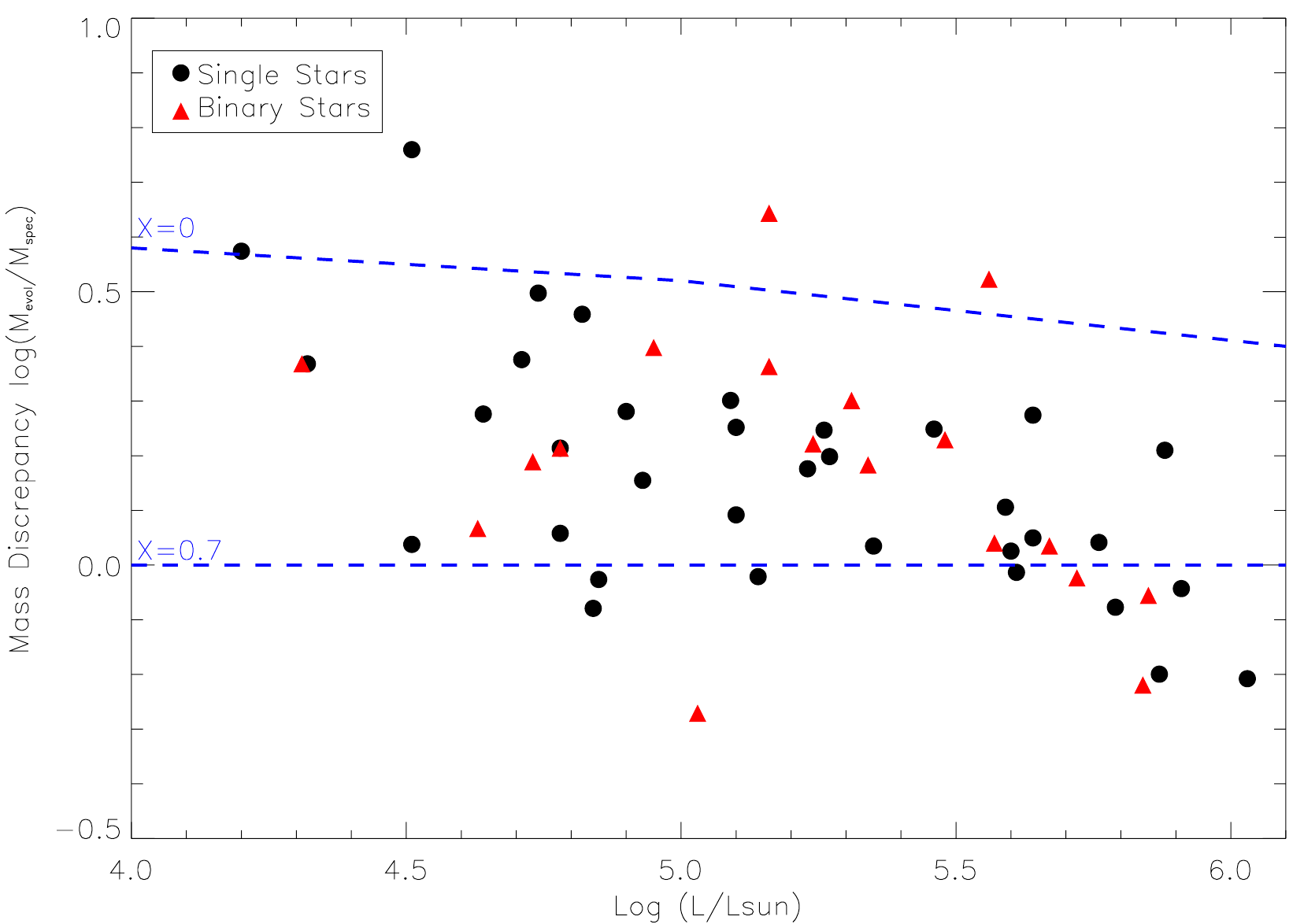,width=\linewidth,angle=0}\\
\caption{Mass-Discrepancy Diagram (i.e. ratio of spectroscopic and evolutionary masses plotted verses luminosity) for the complete supergiant sample. The upper dashed blue line is from the L(M) relation of helium stars: a star of a given mass cannot be more luminous than this. The lower blue line is $M_{evol} = M_{spec}$, and is a lower limit as stars cannot be under luminous.}
\label{f_mass_dis}
\end{figure} 

 \begin{figure}
\epsfig{file=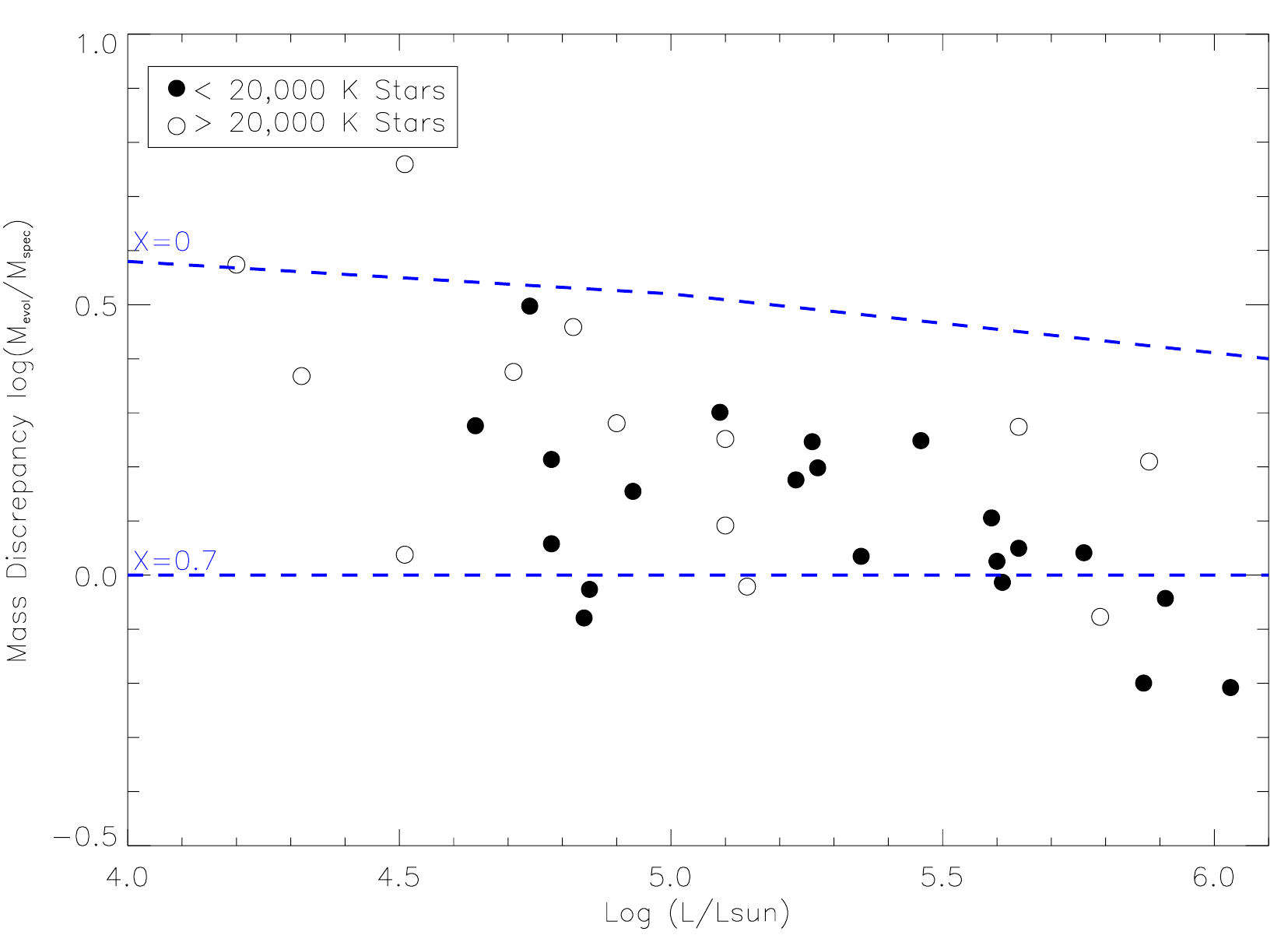,width=\linewidth,angle=0}\\
\caption{Mass discrepancy plotted versus luminosity for only the single stars. Closed symbols are those with effective temperature < 20\,000 K and open symbols are singles with \teff $\ge$ 20\,000 K. The upper dashed blue line is from the L(M) relation of helium stars: a star of a given mass cannot be more luminous than this. The lower blue line is $M_{evol} = M_{spec}$, and is a lower limit as stars cannot be under luminous.}
\label{f_mass_dis_singles}
\end{figure} 

\subsection{Runaway Candidates}

Two from our single-star sample (VFTS\, 165 and VFTS\, 831) have been identified as runaway candidates, in Sana et al. (in prep.) and \cite{eva14}, respectively, because of their unusual radial velocities. Both of these stars have high nitrogen abundances, VFTS\, 165 having a nitrogen abundance derived from the singlet at 3995\AA\, of 7.83 dex and VFTS\, 831 of 7.94 dex. These abundances are large, but still consistent with the rest of the single sample. As these have been identified as runaways it may be possible to explain these higher-end nitrogen abundances using a binary supernova scenario \citep{bla61}, because of the accretion of N-rich gas from a common envelope interaction with a progenitor of a supernova. The positions of these two candidate runaways are marked in Fig. \ref{f_positions}.

\subsection{Projected rotational velocities in single and binary samples}

The estimates of the projected rotational velocities (\vsini) must be considered with some caution as {\em normally} stellar rotation is not the main contributor to line broadening in supergiants. However, given that a consistent methodology was used for both samples, their distributions are worth comparing. In Fig. \ref{cdf}, the cumulative distribution functions (CDFs) are shown for the single and binary samples. These imply that the binary sample have the systematically larger projected rotational velocities. Two-sided Kolmogorov-Smirnov (KS) and Kuiper tests return probabilities (adopting an estimate of half the upper limit for VFTS\,291) of approximately 1\% and 4\% that the samples came from the same parent distribution. 

It is unlikely that these differences would arise from misclassification of targets as single or binary. In such a case, one would expect the broader lined (with larger \vsini\ estimates) binaries to have been classified as single, leading to a difference that would be the reverse of what is observed. Synchronisation of rotational and orbital motion, however, could be important. As an illustrative example, we consider VFTS\,291 which has the smallest gravity (and hence amongst the largest radius) of the binary sample. The equatorial velocity (\ve) will be inversely proportional to the period for circular orbit and implies \ve = 200 \kms for a ten day period. Even a relatively high gravity supergiants such as VFTS\,779 would have \ve = 80 \kms for such a period. The periods of our supergiants are unknown. However our time sampling of their spectra will preferentially have identified short-period binaries and hence the effects of any synchronisation could be significant.

\begin{figure}
\epsfig{file=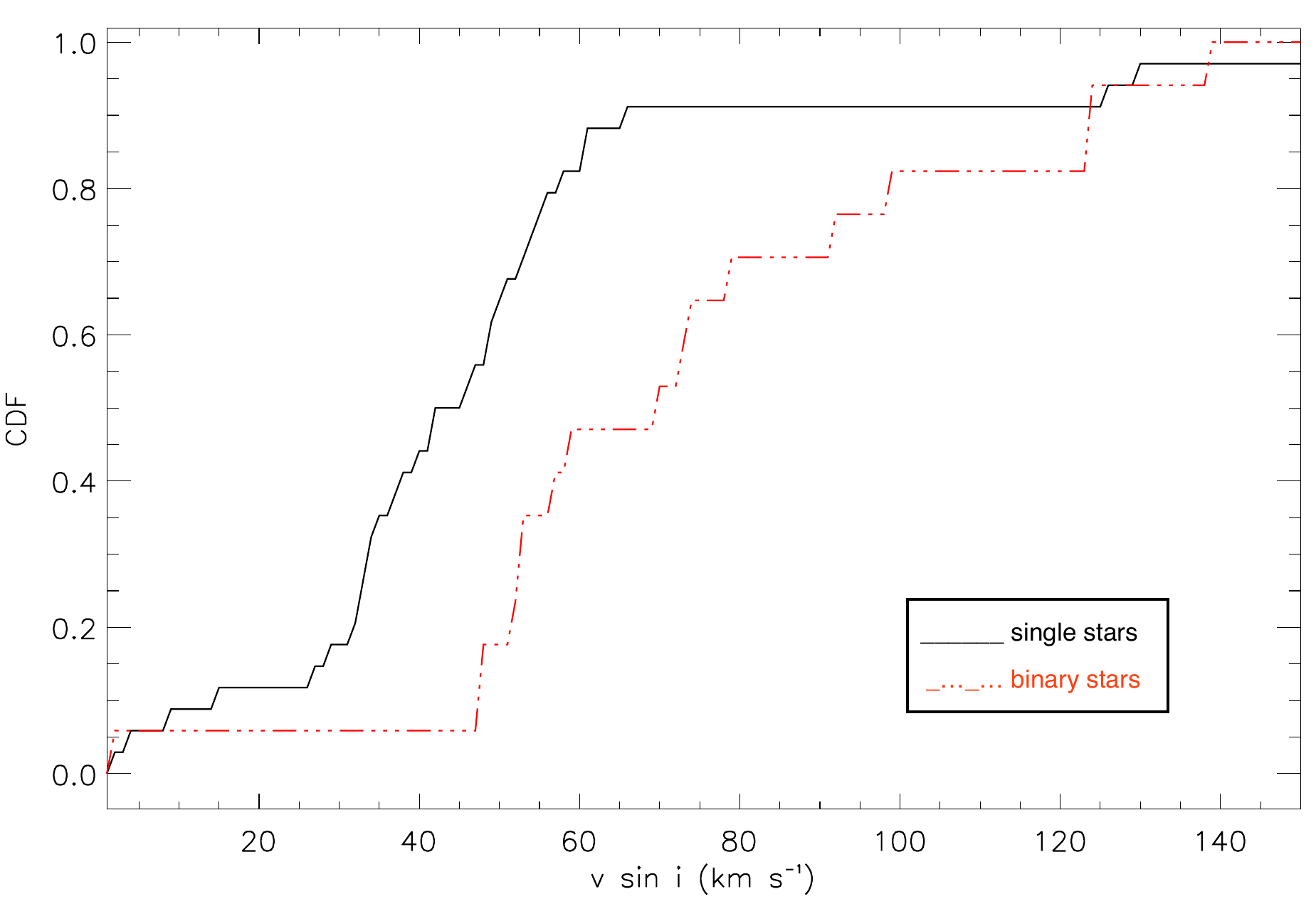,width=\linewidth,angle=0}\\
\caption{Cumulative distribution fucntions (CDFs) for the single (black, solid line) and binary (dotted-dash, red line) supergiant samples as a function of their projected rotational velocities.}
\label{cdf}
\end{figure} 

\subsection{Atmospheric parameters for the single and binary samples}\label{s_atm}

In Fig. \ref{teff_sg}, we investigate the effective temperature estimates as a function of spectral type. The sample has been divided into four groups, single or binary and luminosity classes Ia+/Ia/Iab or Ib/II. Any systematic effective temperature difference with luminosity class would appear to be smaller than the intrinsic scatter at a given spectral sub-type. Additionally, there are no significant differences between the single and binary samples, implying that for the latter the effects of the (normally)  unseen companion would appear to be small.

A simple third-order polynomial fit is also shown in Fig. \ref{teff_sg} and would lead to the effective temperature estimates as a function of spectral type listed in Table \ref{teff_st}. Also listed in this table are the estimates obtained from using a similar procedure on the Galactic samples of \citet{fra10} and \cite{cro06} and the LMC sample of \citet{tru07}. \citet{fra10}, using the same model atmosphere grid as adopted here, analysed the spectra of over 50 Galactic supergiants with luminosity classes ranging from Ia$^+$ to II. \citet{cro06} analysed the spectra of 25 class Ia supergiants using the model atmosphere code {\sc cmfgen} \citep{hil03}. Additionally they considered some other analyses when providing their estimates. \citet{tru07} provided effective temperature estimates for supergiants, giants, and main-sequence stars in the Galaxy and the Magellanic Clouds based on FLAMES spectroscopy; here we list their scale for supergiants in the LMC.

The agreement between the four scales is satisfactory especially considering the different metallicity regimes, luminosity classes and model atmosphere techniques. For example, the absolute differences between the estimates found here and in other studies are typically 500\,K and are never more than 1500\,K. This comparison provides support for the robustness of our techniques and in particular our adoption of a plane-parallel photospheric geometry. This  mirrors  the conclusions of \citet{duf05}, who compared {\sc fastwind} and {\sc tlusty} analyses of supergiants in the SMC and found good agreement.

\begin{figure}
\epsfig{file=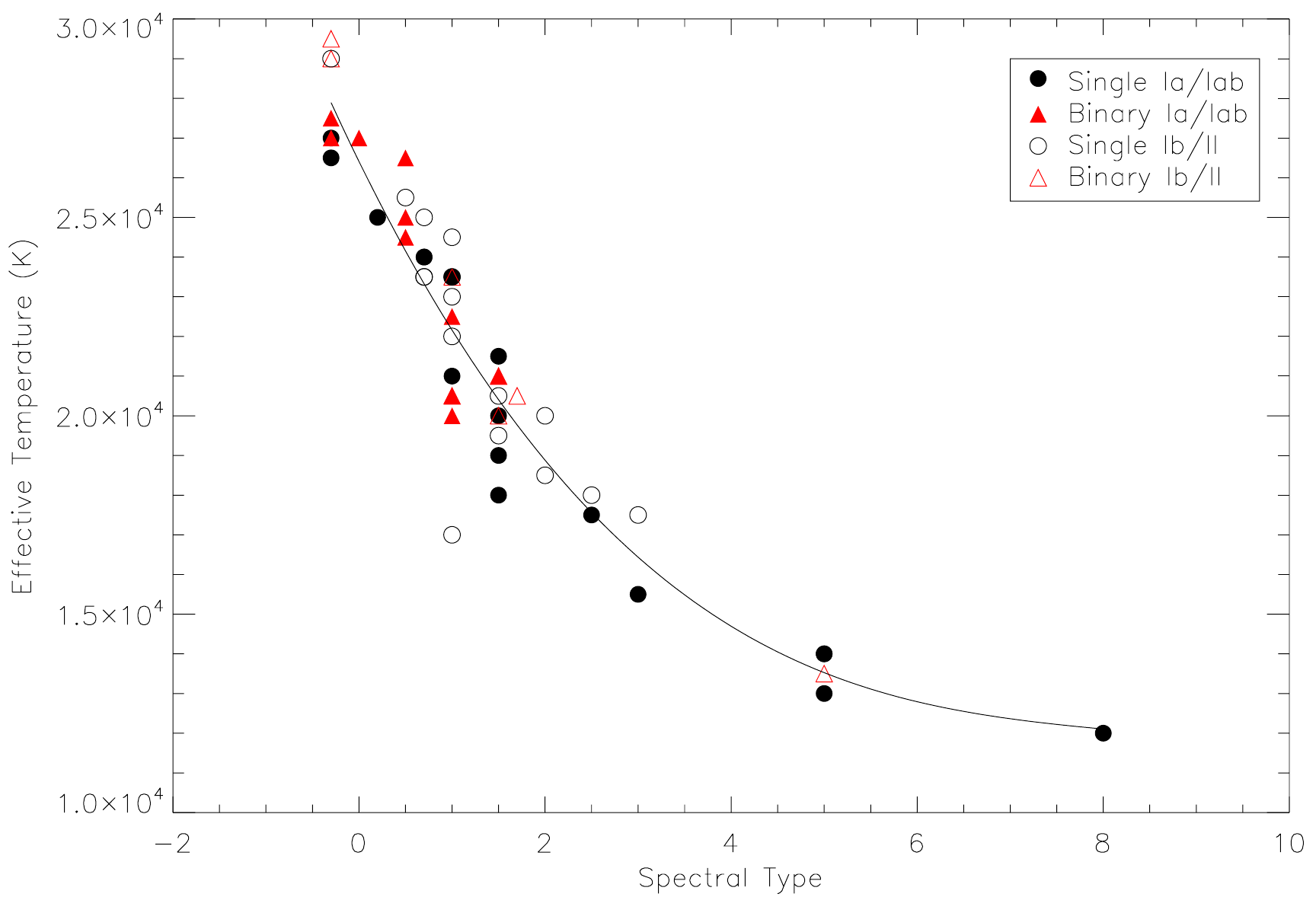,width=\linewidth,angle=0}\\
\caption{Effective temperature estimates plotted as function of B-spectral type (negative values cover O9-B0). The solid line is a best-fit polynomial.}
\label{teff_sg}
\end{figure} 

\begin{table}

\caption{Estimates of effective temperatures (in kK) for different spectral types for the current sample (TAR), for supergiants in the LMC \citep[][LMC]{tru07}, a sample of Galactic supergiants \citep[][GAL]{fra10} and for class Ia supergiants \citep[CLW]{cro06}. 
The LMC sample has been rounded to the nearest 500 K to maintain consistency.}\label{teff_st}
\begin{center}
\begin{tabular}{lrccccccc}
\hline\hline
ST & TAR & LMC  & GAL & CLW 
 \\	\hline

O9.7        &		28.5  &  -     &    -      &    28.5      \\	
B0 		&	        27.0    & 28.5  &  28.0   &    27.5	 \\
B0.5  	&		24.5  & 25.5  &  24.0   &    26.0	   \\
B0.7  	&		23.5  & 23.5  &  23.0   &    22.5	   \\
B1 		&	        22.0   & 22.0  &  21.5   &    21.5	   \\     
B1.5 	&		20.5  & 20.5  &  19.5   &    20.5	   \\
B2 		&	       19.0     & 19.0  & 18.0    &   18.5	   \\   
B2.5  	&	       17.5     & 17.5  & 17.0    &   16.5       \\	
B3  		&	       16.5     & 16.0  & 16.5    &   15.5	   \\
B5		&		13.5	 & -	     &  15.0    &	-	   \\
B8		&		12.0  & -	     &  12.0	   &	-	  \\
\hline
\end{tabular}
\end{center}
\end{table}

The effective temperature estimates listed in Table \ref{t_Atm} and illustrated in Fig. \ref{teff_sg} imply that there is a lack of cool binaries. For example twelve out of the thirty four single stars have \teff $<$ 20\,000K, compared with one out of eighteen of the binaries. However  a Kolmogorov-Smirnov test returned a relatively high probability ($\sim$10\%) that these are drawn from the same parent distributions, with a Kuiper test returning a higher probability. This may reflect the relatively small sample sizes. If the difference was real and the supergiants were evolving from the main sequence, it would require the destruction  of the binaries on relatively short timescales.

The estimated gravities for the single and binary candidates shown in Fig. \ref{f_HRD} are in excellent agreement. Assuming that this reflects the underlying parent populations, it implies that the estimated gravities (and by inference the other estimates) for the binaries have not been significantly affected by contamination of the spectra by the secondary.

\subsection{Element abundances}

Although silicon abundance estimates were obtained, they cannot be considered independent as they were used to constrain the microturbulence. By contrast, those for magnesium are independent. As discussed in Sect. \ref{Ad_ap}, these were estimated as a check of the validity of the analysis and in particular whether they were consistent with the magnesium abundance of the LMC. 

For the 33 single-star estimates, the mean value is 7.02$\pm$0.11 dex, which is consistent with a LMC baseline magnesium abundance of 7.05 dex \citep[see][and references therein]{hun07}. Additionally, the standard deviation is less than the uncertainties inferred in Sect. \ref{Ad_ap}. For the 17 binary supergiants with estimates, the mean value of 7.00$\pm$0.17 dex is again consistent  with the LMC abundance, whilst the increased standard deviation may reflect uncertainties relating to the contamination of the spectra by an unseen secondary. However, for both samples the results are encouraging and imply a typical uncertainty in the magnesium (and by implication the nitrogen) abundance estimates of less than 0.2 dex.

The nitrogen abundance estimates cover a large range from approximately 6.9 dex to 8.3 dex. The former is consistent with the current nitrogen abundance of the interstellar medium in the LMC \citep[see discussion by][and references therein]{hun07}, with the latter corresponding to an increase by approximately a factor of twenty five. 

Carbon abundances are also a useful diagnostic for understanding stellar evolution. For example, an analysis of A-, F-, and G-type Galactic supergiants by \cite{lyu14} found both an underabundance of carbon together with an anti-correlation between the carbon and nitrogen abundances. These were consistent
with mixing between the stellar core and surface during either a main sequence (MS) 
and/or the first dredge-up phase. Unfortunately, in this case, it is not possible to search for a corresponding decrease in the carbon abundance due to the contamination of the \ion{C}{ii} spectra by recombination emission from  \ion{H}{ii} regions. This is illustrated in Fig. \ref{cii}, which shows the spectra region containing the \ion{C}{ii} doublet at 4267\AA. The upper two spectra are for {\em relatively rapidly rotating} near main sequence B-type stars. They contain a narrow emission or absorption feature produced by the subtraction of either two little or too much sky background. From visual inspection, the former would appear to be more common than the latter and may reflect the contribution of a local Str\"omgren sphere to the stellar spectra. 

The bottom spectrum in Fig. \ref{cii} is for the supergiant, VFTS\,315, which has line widths comparable to the spurious features in the upper plots. Hence it would be very difficult (if not impossible) to estimate the degree of contamination of its \ion{C}{ii} feature from an inappropriate correction for the sky background. This uncertainty is endemic to fibre-feed spectroscopy where the sky spectra are spatially separated from the stellar spectra. 

By contrast, the \ion{O}{ii} spectrum is not significantly affected by \ion{H}{ii} region emission and the corresponding  abundance estimates should be reliable. Unfortunately, it only suffers a mild depletion even for significant nitrogen enhancements. For example, an LMC model \citep{bro11a} with a mass of 15\Msun\, and \vor$\sim$400\,\kms achieves a nitrogen enrichment of approximately 1.0 dex before it become a supergiant (here defined as \logg $\sim$3.25 dex). By contrast, the oxygen depletion is less than 0.1 dex and given the observational and theoretical uncertainties, such small variations could not be reliably measured.

A critical examination by \citet{mae14} of the nitrogen abundances estimated in the FLAMES survey \citep{eva05, hun08b} advocated the use of N/C against N/O abundance ratio diagrams. In theory, this could provide a powerful diagnostic. However in practice, it fails to recognise the inherent uncertainties of fibre-feed spectroscopy for the \ion{C}{ii} spectrum. Indeed, such plots are unlikely to be useful for any large scale fibre-fed surveys of young  narrow-lined B-type stars. Hence we have not attempted to use this approach but will instead concentrate on the more limited (and secure) interpretation of the nitrogen abundance estimates. 

\begin{figure}
\epsfig{file=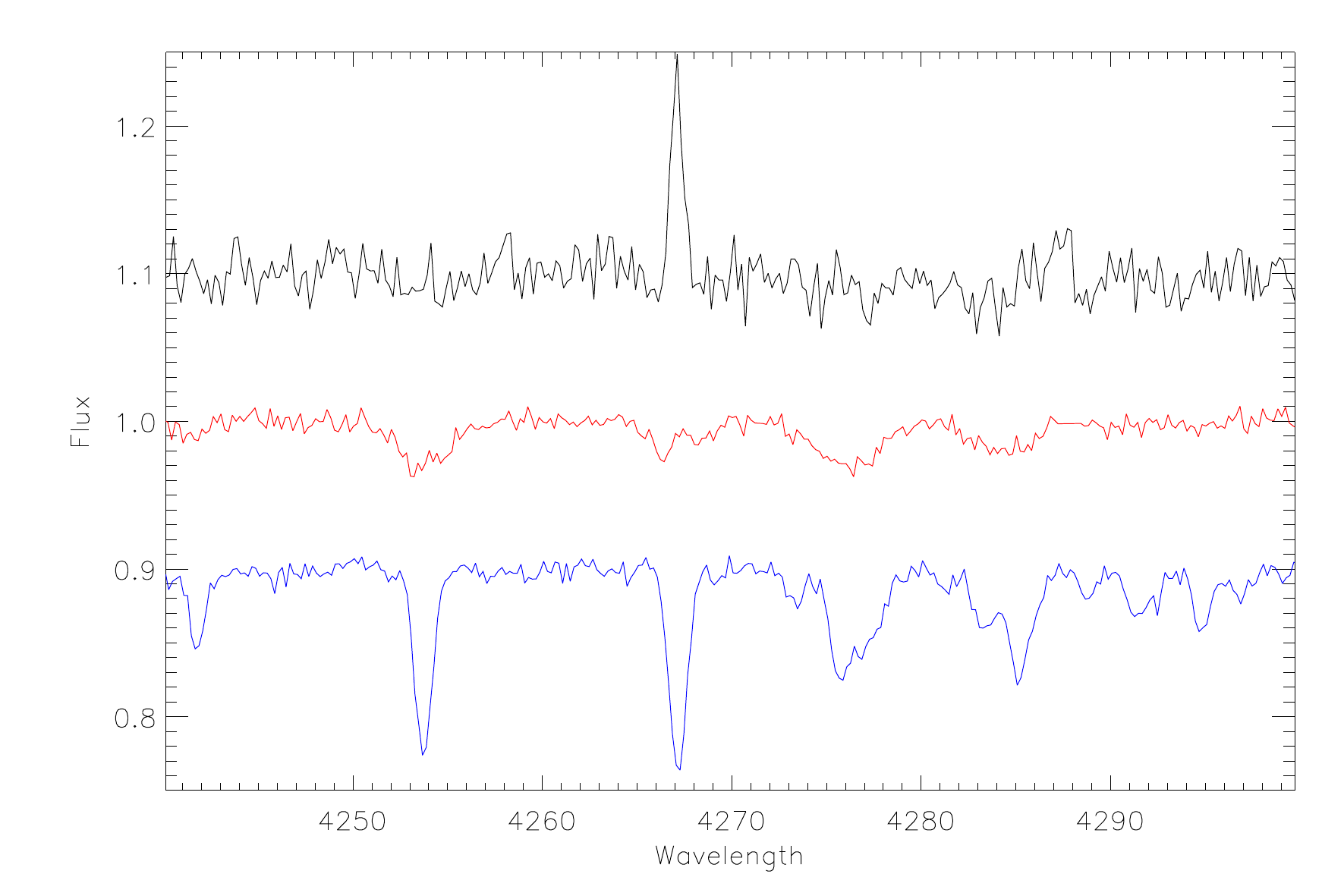,width=\linewidth,angle=0}\\
\caption{Normalised sample spectra encompassing the close \ion{C}{ii}  doublet at 4267\AA. The upper two plots are for the rapidly-rotating near main-sequence stars, VFTS\,535 and 453. The lowest plot is for the supergiant, VFTS\,315. The spectra have been shifted in the y-direction for clarity.}
\label{cii}
\end{figure} 

\subsection{Nitrogen abundances and atmospheric parameters}\label{s_N_atm}

In Fig. \ref{N_atm}, the distribution of nitrogen abundance estimates with both effective temperature and gravity are illustrated. Overlaid are single-star evolutionary models \citep{bro11a} for two sets of initial rotational velocities ($\approx$ 200 \kms and 400 \kms) and four initial masses (12, 20, 40, and 60 \Msun). Models with 0 \kms rotational velocities lie on the black line marking the baseline nitrogen abundance in the LMC. These plots are useful to search both for systematic errors in the abundance estimates and for evidence of evolutionary changes. No systematic trends with gravity are apparent, whilst the distribution of the single and binary samples are similar. 

In the case of effective temperatures, there would appear to be a lack of cool objects with (near) baseline LMC abundances. Although the \ion{N}{ii} spectrum is intrinsically weaker in these objects, abundance estimates (or limits) have been found for the whole sample. Hence the absence of cool supergiants with low nitrogen abundances is unlikely to be due to a selection effect.

The evolutionary models imply that there should be effectively no enrichment as supergiants evolve from higher to lower effective temperatures, for a star of a particular mass. Additionally the nitrogen abundances are in general consistent with the predictions of rotational mixing. For example all enhancements are less than those predicted for \vor $\approx$ 400 \kms. Other recent studies such as those by \cite{lyu14} of A-, F-, and G-type supergiants in our Galaxy, have also shown good agreement with rotational stellar models with initial rotational velocities of 0-300 \kms.

 However there may be some differences between the two samples, e.g. a lack of binaries amongst the cooler supergiants (see Sect. \ref{s_atm}), whilst binaries dominate the sample of stars with nitrogen abundance less than 7.1 dex. In turn this would imply that low nitrogen abundances are more common in binary systems -- we discuss this further in Sect. \ref{s_SvB}.

\begin{figure}
\epsfig{file=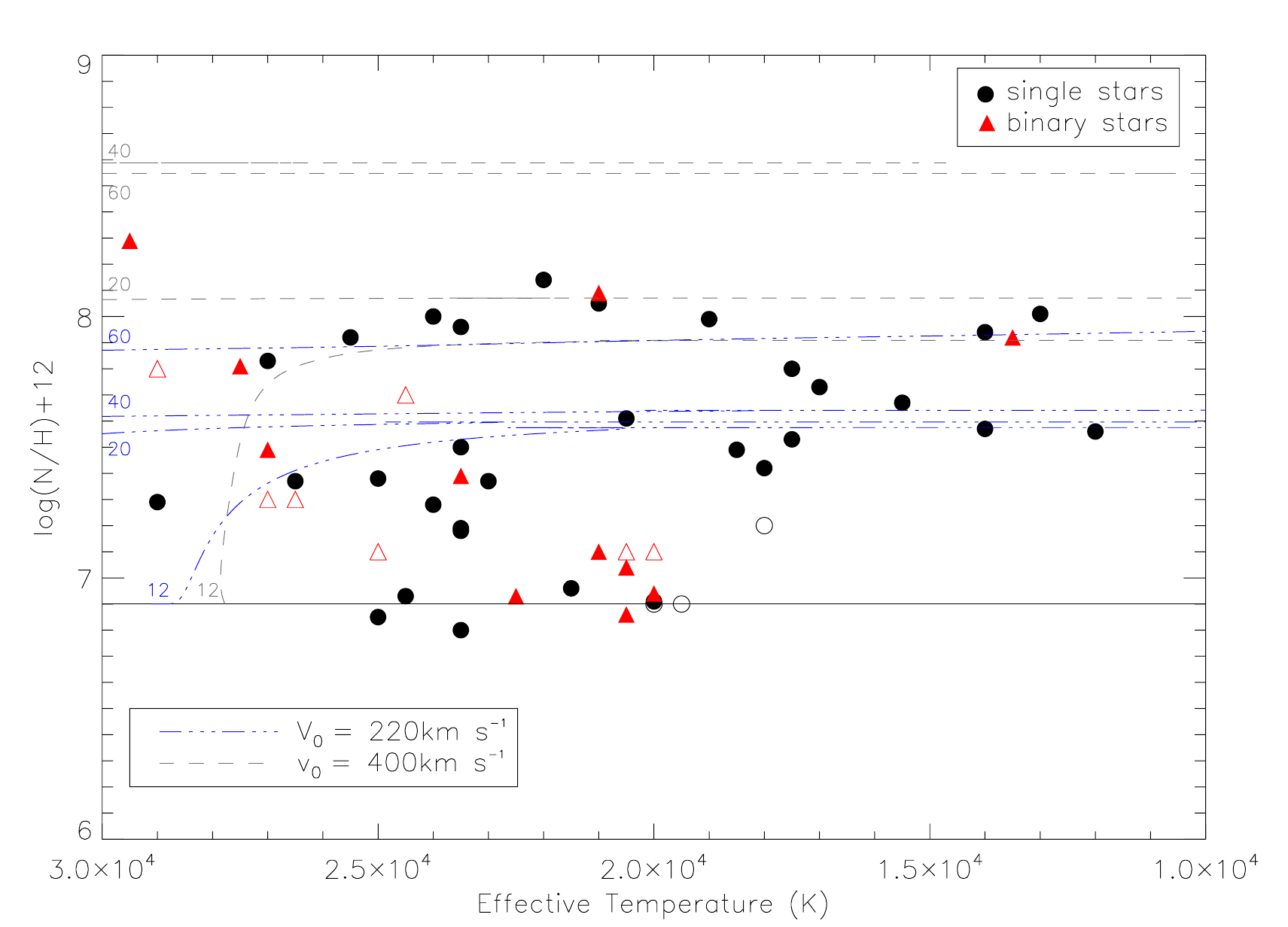,width=\linewidth,angle=0}\\
\epsfig{file=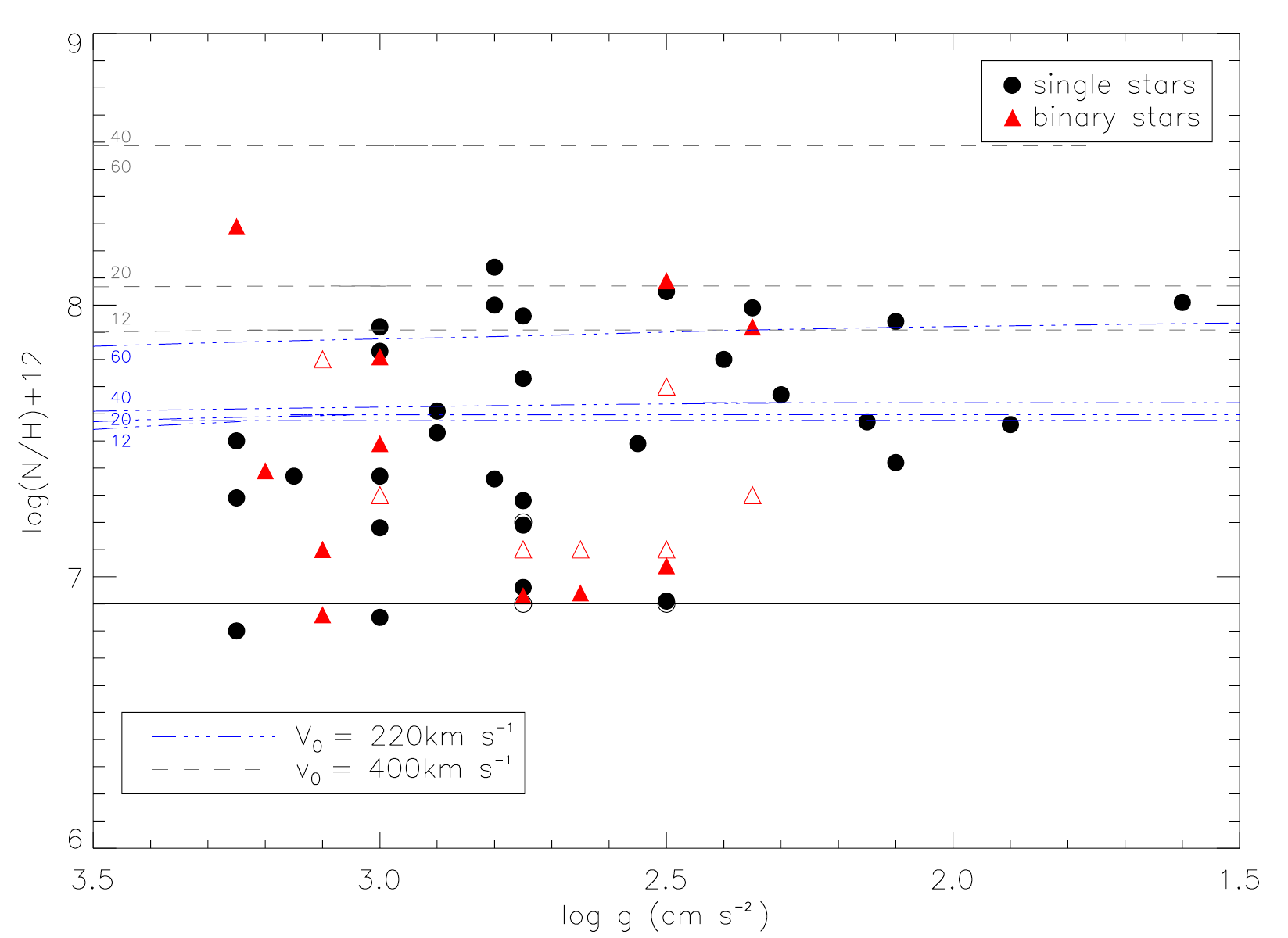,width=\linewidth,angle=0}\\

\caption{Nitrogen abundance estimates as a function of effective temperature ({\emph {upper box}}) and logarithmic gravity  ({\emph {lower box}}). Open symbols represent upper limits for the nitrogen abundance. Over plotted are \cite{bro11a} models at 12, 20,40, and 60 \Msun for two initial rotational velocities of 220\kms (blue) and 400\kms (grey).}
\label{N_atm}
\end{figure} 

\subsection{Nitrogen abundances and spectroscopic masses}\label{s_N_mass}

Nitrogen abundance estimates are plotted against spectroscopic mass estimates for the single and binary supergiants in Fig. \ref{N_mass}. No clear trends with mass are apparent, and the distribution of single and binary stars are similar for stars with spectroscopic mass estimates $\le$~40~\Msun. By contrast, all the higher mass supergiants show significant nitrogen enhancements. However, the majority of our sample have masses $<$~40~\Msun, and hence it would be inappropriate to draw any definite conclusions from the small number of targets in the higher mass range.

\begin{figure}
\epsfig{file=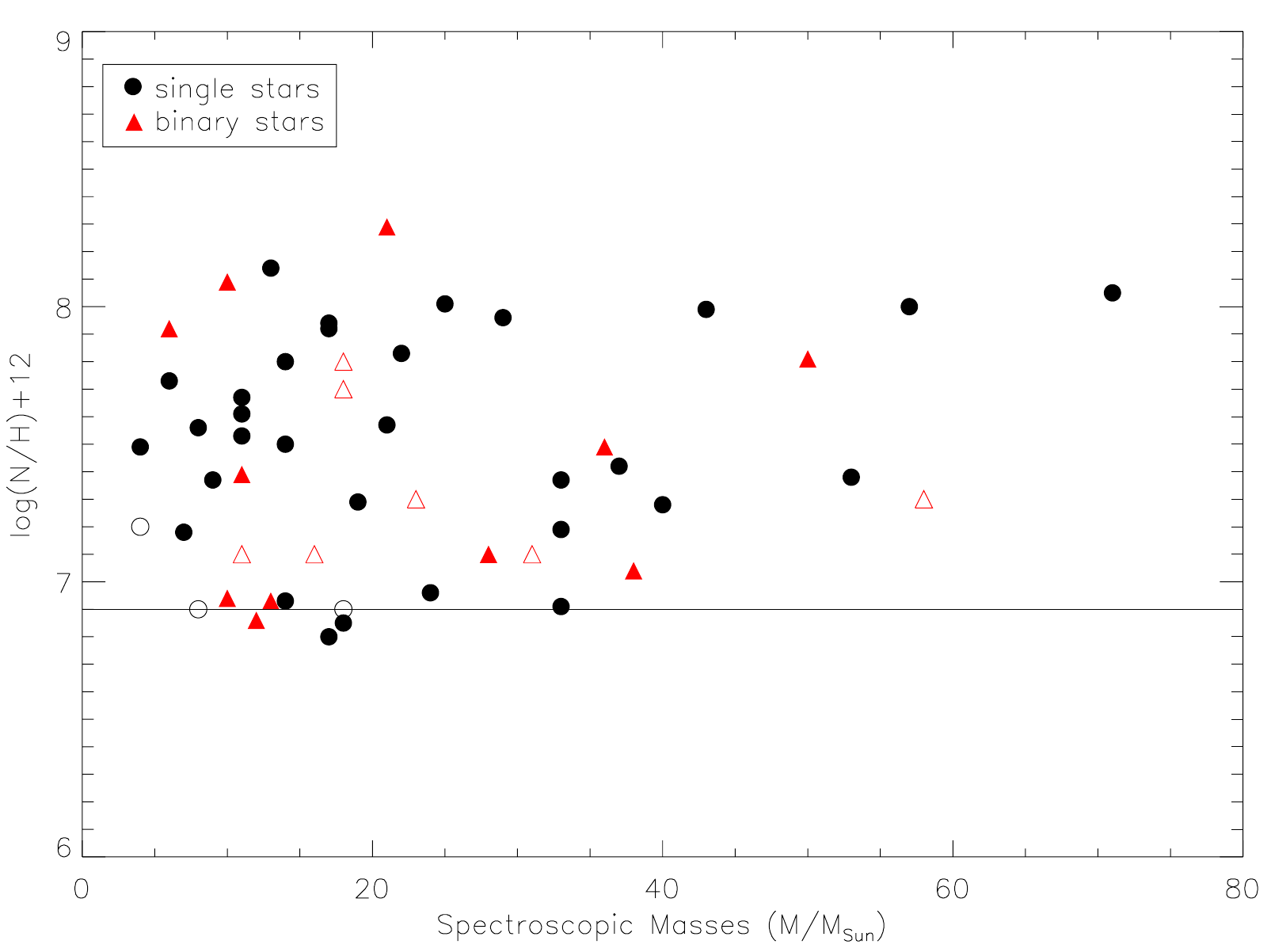,width=\linewidth,angle=0}\\

\caption{Nitrogen abundance estimates as a function of spectroscopic masses. Open symbols represent upper limits for the nitrogen abundance.}
\label{N_mass}
\end{figure}

\subsection{Comparison of nitrogen abundances with morphological classification}

Morphological classifications have been undertaken for all the supergiants as discussed in Sect. \ref{s_obs} and these are listed in Table \ref{t_Targets}. As well as spectral type and luminosity class, some of the supergiants have been classified as either nitrogen strong or nitrogen weak. These classifications have been compared with the nitrogen abundances estimates listed in Table \ref{t_Atm}; for consistency we have used the estimates from the singlet at 3995\AA, although the principal conclusions would remain unchanged if we had used those from all the \ion{N}{ii} features. 

We have divided our sample into three subgroups, viz. nitrogen weak (Nwk/Nwk?), strong (Nstr) and no designation. The nitrogen weak sample contains 15 targets, with five targets - all of which are binary candidates - having only upper limits (ranging from 7.1 to 7.7 dex). This prevalence of upper limits in the binary sample may reflect the slightly lower S/N of their spectra, their higher effective temperatures or an intrinsically lower nitrogen abundance. We consider this further in Sect. \ref{s_SvB}. The remaining 10 nitrogen weak supergiants have nitrogen abundance estimates in the range 6.9-7.2 dex with a mean value of 7.02$\pm$0.14 dex. 

By contrast, there are only two nitrogen strong supergiants (one single, one binary) and they have abundance estimates of 7.99 and 7.81 dex respectively. Finally, there are 34 normal supergiants (of which nine are binary candidates). Five of these targets have upper limits for their nitrogen abundances with the remainder having a range from 6.8 to 8.14 dex and a mean value of 7.64$\pm$0.45 dex.

It is encouraging that there is a good correlation between the mean nitrogen abundances and their morphological classification. Those without a designation have a relative wide range of abundances with a mean nitrogen enhancement of a factor of six over the baseline LMC nitrogen abundance of 6.9 dex \citep[see][and references therein]{hun07}. By contrast, the nitrogen weak group have a relatively small spread of abundances with enhancements being a factor of two or less. Indeed with the uncertainties discussed in Sect. \ref{Ad_ap}, all these targets could have the baseline LMC nitrogen abundance. By contrast, the (small) group of nitrogen strong targets, both have significant nitrogen enhancements with a mean enhancement of a factor of ten. These results are in accordance with the hypothesis of \cite{wal76} that OBC (and N-weak) supergiants have normal (main-sequence) abundances, the morphologically normal majority are nitrogen enhanced, with the OBN (and N strong) being even more so.

\subsection{Nitrogen abundances in the single and binary samples}\label{s_SvB}

One of the principal aims of this analysis was to search for evidence of different evolutionary histories within the single and binary supergiants. Such a search will be complicated by the possibility that some of the single supergiants may {\em currently} be an undetected binary or may have evolved from a binary system \citep[see, for example,][]{deM14}. For the former, undetected binaries will be weighted towards those having compact companions or long-period systems \citep{san13}, which may have effectively evolved as single stars.

From Table \ref{t_Atm}, it would appear that the distributions of nitrogen abundance within the single and binary samples are broadly similar with ranges of approximately 6.9 - 8.1  and 6.9 -- 8.3 dex, respectively. The mean and standard deviations excluding stars with upper limits are 7.52$\pm$0.39 and 7.44$\pm$0.51 dex, whilst including stars with upper limits (with arbitrary estimates midway between the limit and the LMC baseline value) leads to 7.47$\pm$0.41 and 7.32$\pm$0.43 dex, respectively. Additionally the analysis in Sect. \ref{secondary} implies that the nitrogen abundances for the binary sample may be slightly underestimated. 

In Sect. \ref{s_N_atm} and Fig. \ref{N_atm}, some tentative evidence was found for low nitrogen abundance targets to be more prevalent in binary systems. To further investigate this we have created CDFs for the nitrogen abundances of single and binary targets (using the prescription given above to deal with upper limits) and these are illustrated in Fig. \ref{cdf_N}. 

\begin{figure}
\epsfig{file=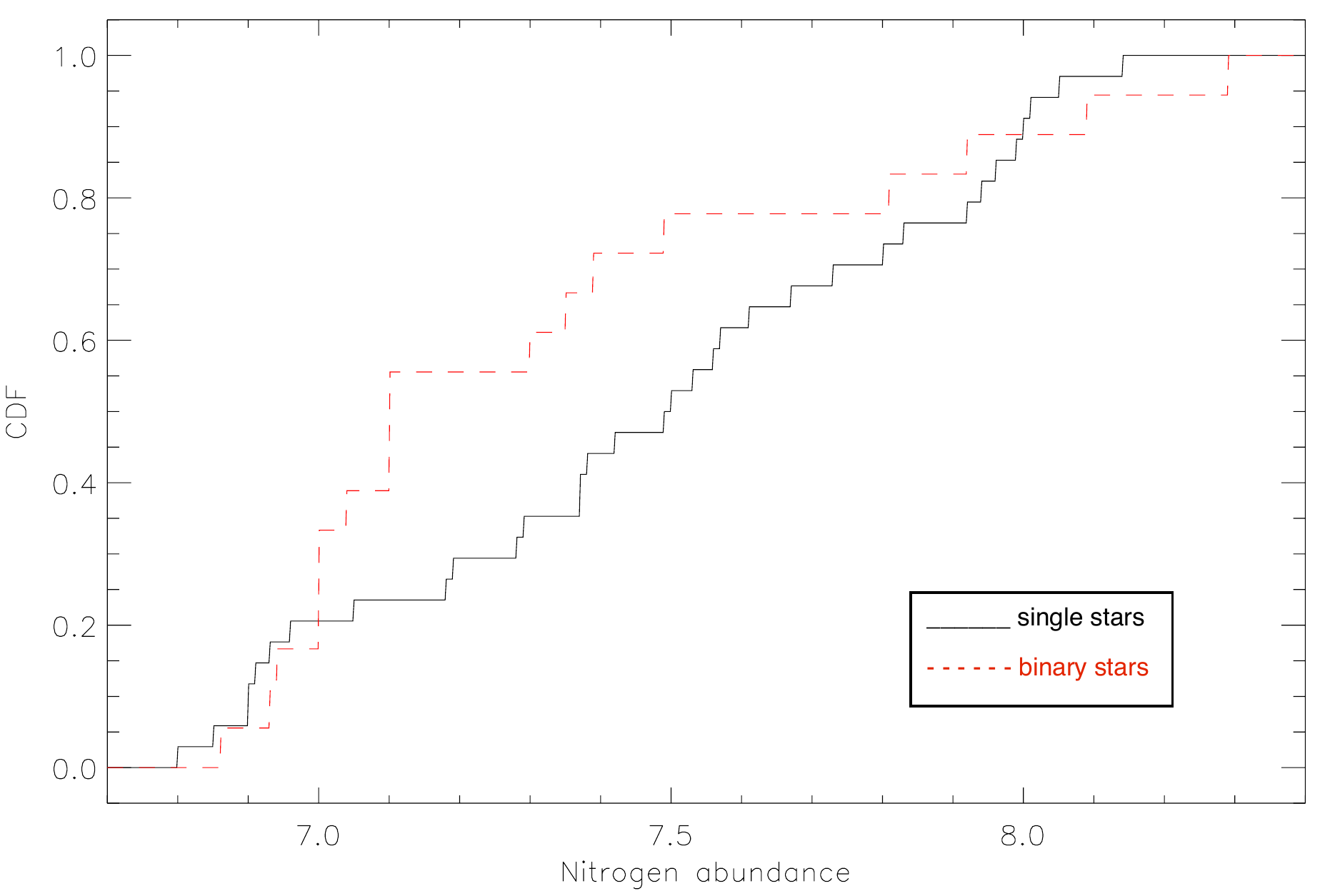,width=\linewidth,angle=0}

\caption{Cumulative distribution functions for nitrogen abundance estimates for the single (solid line) and binary (dotted line) supergiants.}
\label{cdf_N}
\end{figure} 

Two sided Kolmogorov-Smirnov (KS) and Kuiper tests returned similar probabilities of approximately 15\% that the samples were drawn from the same parent population. Although this indicates that they may come from different populations, it is not in any way definitive. However, if the differences were real then this would imply that lower nitrogen abundances were more prevalent in the binary sample, which also contained some highly-enriched primaries. The former would be consistent with an (ongoing) analysis of slowly rotating B-type {\em near main sequence} stars in the VFTS survey. Preliminary results indicate that the binaries generally do not show the large (\ga 0.5 dex)  nitrogen enrichment found in some of the single stars. The binary supergiants with low nitrogen abundances would then have evolved from such a binary population assuming that it extended to higher (initial) effective temperatures. By contrast, for the highly nitrogen enriched systems, there would appear to have been significant interaction between the components \citep[see, for example,][for a discussion of VFTS\,450 and 652]{how15}. 

\subsection{Nitrogen abundances and projected rotational velocities}

The nitrogen abundance estimates (from the \ion{N}{ii} singlet at 3995\AA) are shown as a function of projected rotational velocity for the single and binary supergiants in the upper panel of Fig. \ref{N_vsini}.

%****************************************************************************************************************

To increase the sample size, we have also included comparable results from the FLAMES-I survey taken mainly from \citet{hun07} and \citet{tru07}\footnote{See http://star.pst.qub.ac.uk/sjs/flames/Flames\_Bstar\_data\_table.txt}. For three targets additional nitrogen abundance estimates have been estimated (N11-076: 7.90 dex; NGC2004-005: 7.82 dex; NGC2004-007: 7.73 dex). The same selection criteria were used as for the VFTS sample, viz. that the star had been classified as luminosity classes I or II (there were no luminosity classes II/III) with an estimated gravity, \logg$\leq 3.2$\, dex. The lower panel of Fig. \ref{N_vsini} shows a corresponding plot for the FLAMES-I data with the two targets with upper limits on the nitrogen abundance being excluded. 

Care must be taken when comparing the results from the two surveys as there were significant differences. The VFTS concentrated on one relatively young LMC cluster with the observational strategy tailored to search for binarity. By contrast, the FLAMES-I survey took observations towards two clusters in the LMC (N11 and NGC2004). Normally the exposures did not have the time cadence to reliably distinguish binarity (consistent with the smaller fraction of identified binaries) whilst the majority of the targets were identified as field stars. 

However both datasets show similar features with the majority ($~80-90$\%) of targets having  small projected rotational velocities coupled with a wide range of nitrogen abundances. Additionally, a small number of relatively faster rotating targets exist in both surveys and surprisingly these show no evidence for enhanced nitrogen abundances. In Table \ref{t_high}, we summarise the properties of these targets, with the estimates for the FLAMES-I targets being taken from \citet{hun07} and \citet{tru07}\footnotemark[\value{footnote}].

\begin{figure}
\begin{center}
\epsfig{file=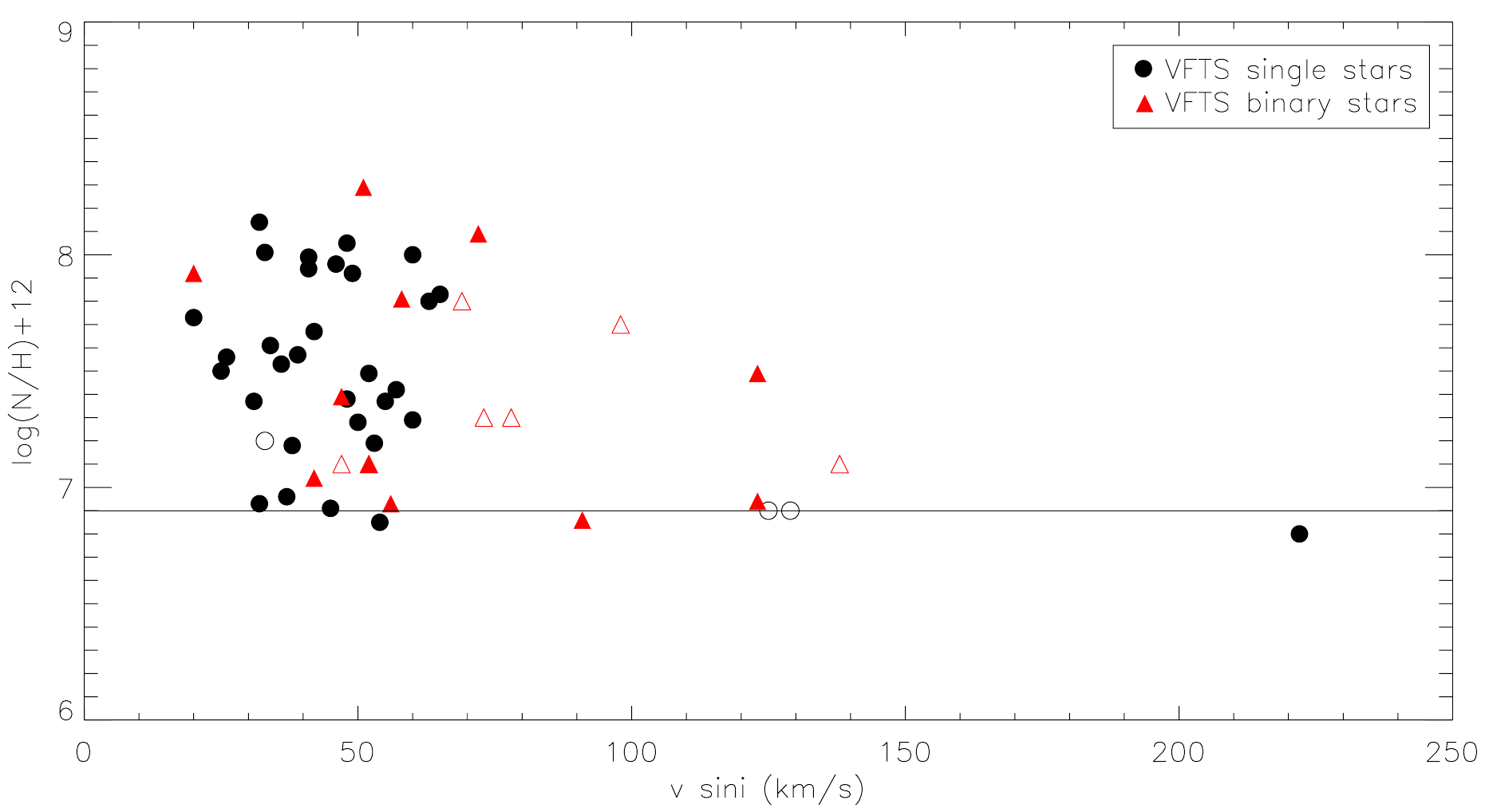,width=8cm,angle=0}\\
\epsfig{file=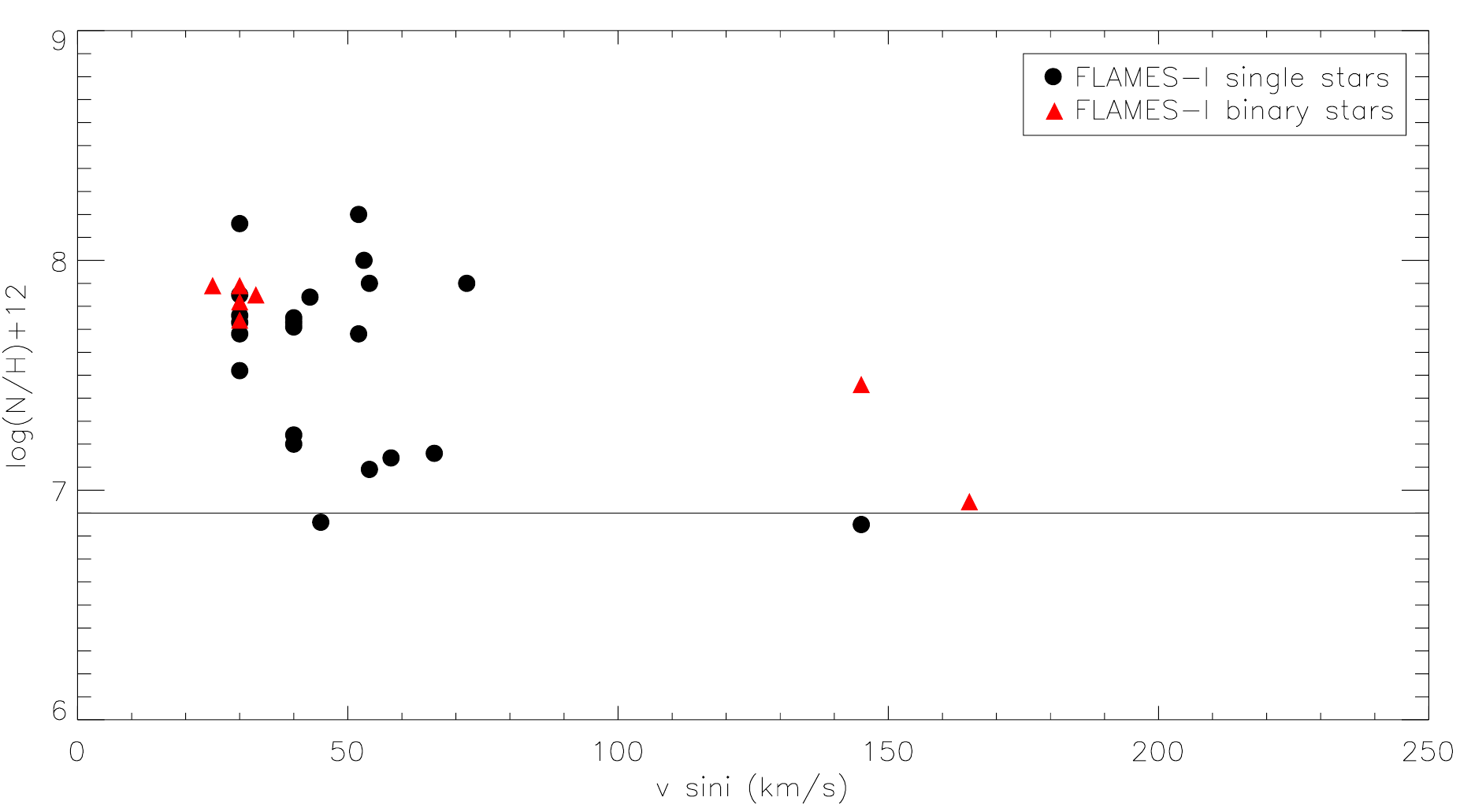,width=8.1cm,angle=0}
\end{center}

\caption{Nitrogen abundance estimates as a function of projected rotational velocity. Estimates are shown for targets in the VFTS ({\em upper box}) and the FLAMES-I survey ({\em lower box}). Open symbols represent upper limits for the nitrogen abundance.}
\label{N_vsini}
\end{figure} 

\begin{table*}

\caption{Properties of supergiants with a projected rotational velocity of more than 100 \kms. The VFTS targets have been taken from Tables \ref{t_Atm} and \ref{t_obs}. The FLAMES-I estimates (identified using the associated cluster, either N11 or NGC2004) have been taken from \citet{hun07} and \citet{tru07} or have been estimated using the same methodology as for the Tarantula targets.  The luminosities and masses are in units of the solar value. The single stellar candidates are listed first followed by the binaries.}\label{t_high}

\begin{center}
\begin{tabular}{lllllrclll}
\hline\hline
Star &  ST & \vsini	& \teff  &	 \logg  & \NtoH &  $\log$\ L/\Lsun &   \Msp/\Msun &    \Mev/\Msun  & 	 $\log t$ 
 \\
	&  	  & \kms   &	K & cm s$^{-2}$&	&	&	   \\	\hline
VFTS\,060	& B1.5 II-Ib((n))			&129	& 19\,500		&2.65	&	$\le$6.9 	& 4.82	& 8	& 23		& 6.84 	\\
VFTS\,293	& B2 III-II(n)e				&125	& 20\,000		&3.00	&	$\le$6.9 	& 4.84 	& 16	& 15		& 7.08	\\
VFTS\,826 	& B1 IIn					&222	& 23\,500		&3.25	&     	6.80       	& 4.85 	& 15	& 16		& 7.04	\\
NGC2004-13  	& B2 II					&145	& 21\,500		&3.20	&	6.85		& 4.94 	& 24	& 14		& 7.13\\
\hline
VFTS\,450	& O9.7 III: +O7::		&123	& 27\,000		&3.00	&	7.49		& 5.67 	& 33 & 39		& 6.59	\\
VFTS\,687	& B1.5 Ib((n)) Nwk			&123	& 20\,000		&2.65	&	6.94		& 4.95 	& 9   & 25		& 6.80	\\
VFTS\,829	& B1.5-2 II				&138	& 20\,500		&2.90	&	$\le$7.1 	& 4.78 	& 10 & 18		& 6.96	\\
NGC2004-15	& B1.5 II					&165	& 23\,000		&3.15	&	6.95		& 4.96	& 22	& 17		& 6.98	\\
NGC2004-20	& B1.5 II					&145	& 23\,000		&3.20	&	7.46		& 4.91	& 25	& 16 	& 7.01	\\
\hline
\end{tabular}
\end{center}
\end{table*}

Excluding VFTS\,450 \citep[which is a peculiar short period SB2 system discussed by][]{how15}, these rapid rotators have relatively low luminosities and large gravities. However in both surveys, there are supergiants with similar physical parameters but with low projected rotational velocities; some of these targets (e.g. VFTS\,307, 845, 855) showed significant nitrogen enhancements.  By contrast, all the higher luminosity (Ia/Iab) supergiants have low projected rotational velocities.

To investigate whether the supergiants with large projected rotational velocities (listed in Table \ref{t_high}) have significantly lower nitrogen abundances, we undertook the following procedure. All supergiants with luminosity classes Ib, Ib-II, and II in both surveys were divided into two samples based using a threshold projected rotational velocity of 100 \kms. The two sample Kolmogorov-Smirnov  and Kuiper statistics that these arise from the same population were 0.0012 and 0.087, respectively. Hence it would appear that the fast and slow rotating cohorts may arise from different populations.

\begin{figure}
\begin{center}

 \includegraphics[width=\columnwidth]{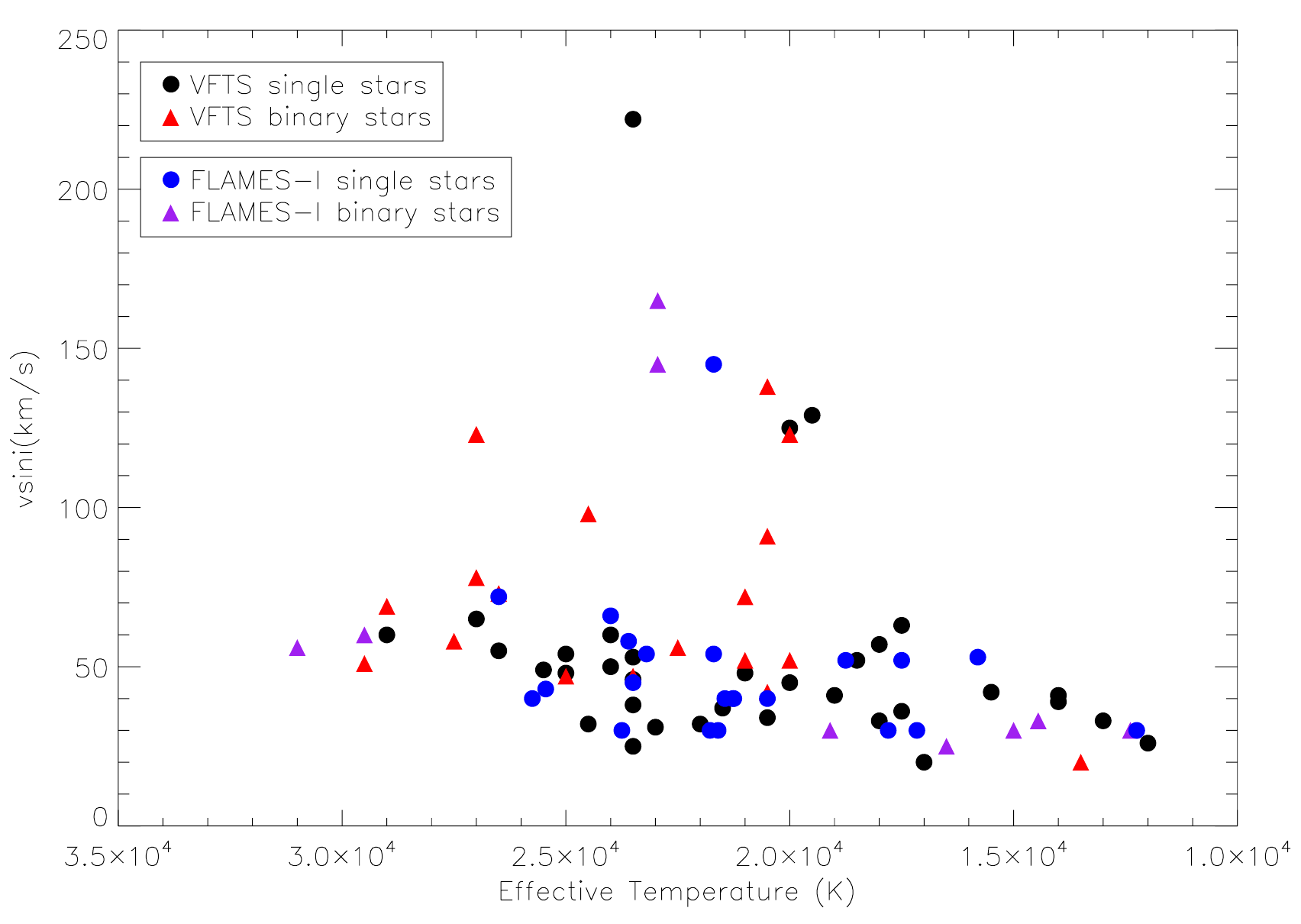}
\end{center}

\caption{Effective temperatures plotted against \vsini\ for FLAMES--I and VFTS supergiants in the LMC.}
\label{teff_vsini}
\end{figure} 

\subsection{Projected rotational velocities and effective temperature}\label{vsini_t}

Figure\,\ref{teff_vsini} shows projected rotational velocity plotted versus 
the effective temperature for all blue supergiants in the LMC from 
the original FLAMES--I survey \citep{hun07, tru07} and the current VFTS data. 
In this plot, a steep drop in \vsini\ can be seen at $\sim$19\,500 K (from as high as 220 \kms for higher temperature stars to <60\kms \,for all cooler stars). 
A similar jump, with a notable absence of rapidly rotating B supergiants, has been seen in other datasets \citep[e.g.][]{how97} as discussed in \cite{vin10}. 

\citet{vin10} provided two possible explanations for this behaviour. 
It could possibly be caused by the bi-stability jump \citep{pau90}, a theoretically predicted discontinuity 
where wind properties change from fast to slow winds, with increased mass loss, when the effective temperature drops below 21\,000K. 
\cite{vin08} suggest that this temperature and that of the steep drop of \vsini\ in observational data \citep[e.g.][]{how97,hun07, tru07} were similar and thus proposed 
bi-stability braking (BSB), i.e. a drop in the surface rotation rates due to increased mass-loss rates, as an explanation. 
Although our drop occurs at 19\,500 K, this process may take some time and so our limit of 19\,500 K seems reasonably consistent with that of the bi-stability jump of 
21\,000K.  This explanation would imply that all the objects, both cooler and hotter than 21\,000K have directly evolved from the main sequence but that a considerable amount of angular momentum is lost as a result of stellar winds. 

The second possibility is that there are two different populations -- one of rapidly rotating stars evolving directly from the zero age main-sequence and another involving 
slowly rotating evolved stars that have lost angular momentum, possibly because they underwent binary evolution or because they are blue-loop, post-RSG stars. 
In any case they would no longer be core hydrogen burning, and they would thus be post main-sequence stars. 

It can be seen from the HRD in Fig. \ref{f_HRD} that at all effective temperatures there is a range of masses, implying that there is no clear bias of only low-mass stars at lower temperatures. In the models of \cite{vin10} BSB can only be used to explain the steep drop for objects above a certain critical mass, but the authors state that if the main sequence were extended sufficiently far to the red (e.g. because of 
enhanced core overshooting), then BSB could account for the drop for stars with masses as low as 15\Msun. This would also help to explain the very existence of blue supergiants, and it is consistent with our discussion in relation to the TAMS in Sect. \ref{ss_HRD}.

\section{Conclusions}

Spectroscopy of the late O- and B-type supergiants observed by the VLT-FLAMES Tarantula survey has been analysed. For the first time, targets showing no significant  radial velocity variations and binary candidates have been considered, in order  to investigate the role of binarity in their evolutionary history.

TLUSTY non-LTE model atmospheres calculations have been used to determine atmospheric parameters and nitrogen abundances for 34 single and 18 binary supergiants.  Effective temperatures were deduced using the silicon balance technique, complemented by the helium ionisation in the hotter spectra. Surface gravities were estimated using Balmer line profiles and microturbulent velocities deduced using the silicon spectrum. Nitrogen abundances or upper limits were estimated from the \ion{N}{ii} spectrum. The effects of a  flux contribution from an unseen secondary  were considered for the binary sample, whilst the validity of the adopted approach was tested by re-analysing some targets using the unified {\sc fastwind} code.

The principal conclusions are:

\begin{enumerate}

\item A comparative study of the single and binary sample shows no major differences between their atmospheric parameters and nitrogen abundances.
\item Sixteen targets are close to, but at lower effective temperatures than the predicted TAMS. This may imply that the TAMS should be moved to slightly lower effective temperatures.
\item The binary sample has fewer cooler objects compared to the single stars, but this may reflect the small sample sizes.  
\item Single-star evolutionary models, which include moderate initial rotation (\vor $\leq$ 400 \kms), can account for all of the nitrogen enhancements observed in both the single and binary samples. The detailed distribution of nitrogen abundances in the single and binary samples may be different, possibly reflecting differences in their evolutionary history. 
\item For the supergiants in the Tarantula survey  and in a previous FLAMES survey, the majority have small projected rotational velocities. In this survey, the distribution peaks at about 50 \kms with 65\% in the range 30 \kms $\ge$ \vsini\ $\le$ 60 \kms. About ten per cent have larger values (\vsini $\ge$ 100 \kms) but surprisingly these show little or no nitrogen enhancement. 
\item All the cooler supergiants have low projected rotational velocities and relatively high nitrogen abundance estimates consistent with  either bi-stability braking or evolution on a blue loop being possible explanations.

\end{enumerate}

This and other samples of Galactic and Magellanic Cloud supergiants now contain a significant number of objects. As such, a more sophisticated analysis of their evolutionary status using, for example, the {\sc bonnsai} stellar evolutionary tools \citep{sch14} should yield further insights.

\begin{acknowledgements}
Based on observations at the European Southern Observatory Very Large Telescope in programme 182.D-0222.
CM is grateful to the Department of Education and Learning (DEL) in Northern Ireland and Queen's University Belfast for the award of a research studentship.
SS-D and FN acknowledge funding from the Spanish Government Ministerio de Economia y Competitividad (MINECO), FN through grants AYA2010-21697-C05-01, FIS2012-39162-C06-01 and ESP2013-47809-C3-1-R and SS-D through grants AYA2010-21697-C05-04, AYA2012-39364-C02-01 and Severo Ochoa SEV-2011-0187, and the Canary Islands Government under grant PID2010119. NRW acknowledges support provided by NASA through grant GO-12179.01 from STScI, which is operated by AURA, Inc., under NASA contract NAS5-26555.
SdM acknowledges support for this work by NASA through an Einstein Fellowship grant, PF3-140105
\end{acknowledgements}

\bibliography{literature}

\end{document}